# Interface-Controlled Defect Engineering in TiN/TaN Superlattices for Enhanced Hardness and Fracture Toughness

Zecui Gao[a,b,c], Qimin Wang[c], Julian Buchinger[a], Nikola Koutna[a], Marcus Hans[d], Zaoli Zhang[e], Jochen Schneider[d], Daniel Primetzhofer[f], Paul Heinz Mayrhofer[a,*] a Institute of Materials Science and Technology, Technische Universität Wien, Getreidemarkt 9, 1060 Vienna, Austria

b Plansee (Shanghai) High Performance Material Ltd., China

c School of Electromechanical Engineering, Guangdong University of Technology, Guangdong, China

d Materials Chemistry, RWTH Aachen University, Kopernikusstr. 10, 52074 Aachen, Germany

e Erich Schmidt Institute of Materials Science, Austrian Academy of Sciences, Jahnstr. 12, 8700 Leoben, Austria

f Materials Physics, Uppsala University, Regementsvägen 10, 752 37 Uppsala, Sweden

*Corresponding author: paul.mayrhofer@tuwien.ac.at

## Abstract

TiN/TaN superlattice coatings were designed to investigate how atomic-scale interface chemistry and defect-stabilized TaN layers govern hardness and fracture toughness. Guided by first-principles predictions identifying TaN-based layers as more damage tolerant than TiN, coherent superlattices with a bilayer period of ~6 nm were synthesized by reactive magnetron sputtering and interfacially doped with C, B, or Si. Structural and chemical analyses reveal coherent fcc architectures with well-defined interfaces. Si segregates preferentially to the interfaces while incorporating into both TiN and TaN, whereas C and B predominantly diffuse into the TaN layers, modifying coherency strain, bonding, and defect populations. Consequently, hardness increases from ~34 GPa for the undoped superlattice to ~41 GPa for the Si-doped architecture, whereas fracture toughness increases from ~2.8 to ~4.0 MPa√m for the B-doped superlattice. First-principles calculations show that vacancy-stabilized $Ta_xN_y$ enhances elastic compliance and elastic contrast rather than intrinsic toughness, while the additional toughening induced by B indicates localized defect-assisted energy dissipation at chemically engineered interfaces. Thus, Si maximizes interface strengthening, whereas B provides the most favourable hardness–toughness balance while preserving high hardness (~38 GPa). These findings establish interface chemistry as an additional design parameter for tailoring the mechanical performance of ceramic nitride superlattices.

# 1 Introduction

Ceramic transition metal nitrides, including TiN, CrN, and (Ti,Al)N, combine high hardness, thermal stability, and chemical resistance, which has established them as indispensable materials for cutting tools, wear-resistant components, and protective coatings in extreme environments [1,2,3,4]. Despite these advantages, their intrinsically limited fracture toughness severely restricts performance under impact loading, thermal cycling, and complex stress states, where crack initiation and propagation govern failure [1]. Overcoming this inherent brittleness requires materials design strategies that extend beyond conventional monolithic coating architectures.

Superlattice (SL) coatings, composed of alternating nanometre-thick layers of distinct materials, provide a powerful approach to address this limitation. By introducing periodic interfaces, superlattices exploit coherency strain, elastic mismatch, and interface chemistry to impede crack propagation and enhance energy dissipation. Improvements in hardness and toughness arise from mechanisms such as crack deflection, interface shearing, and stress delocalization. Crack deflection increases fracture energy by forcing cracks to repeatedly change direction at interfaces, while interface shearing absorbs energy through localized plastic deformation. Stress delocalization further mitigates local stress concentrations that would otherwise promote brittle failure [1,5,6].

Historically, the terms “multilayer” and “superlattice” were often used interchangeably, although they imply distinct structural characteristics. Multilayers refer to alternating layers with controlled thickness, whereas superlattices are defined by coherent or semi-coherent interfaces and a well-defined periodic modulation of lattice and chemistry as well as elastic properties [1,7]. Superlattice architectures have played a central role in the development of hard coatings, providing a unique pathway to overcome the classical trade-off between strength and ductility. As reviewed comprehensively from both historical and mechanistic perspectives in Ref. [1], early concepts of superlattice strengthening evolved from crystallographic ordering in metallic alloys to artificially layered thin films, where coherency strain, elastic contrast, and interface-controlled defect processes govern mechanical response. In nitride-based hard coatings, these concepts enabled hardness levels far exceeding those of the individual constituents, most prominently demonstrated for TiN/VN and related transition-metal nitride superlattices [8].

While the superlattice effect on hardness is well established and can be rationalized within the Koehler–Chu–Barnett framework—encompassing modulus-difference (Koehler [9]) strengthening, coherency-stress hardening at coherent interfaces, epitaxial stabilization of metastable phases, Orowan-like dislocation bowing within individual layers, and, in polycrystalline architectures, additional Hall–Petch strengthening—the simultaneous enhancement of hardness and fracture toughness remains far more challenging, particularly in ceramic-like systems dominated by strong directional covalent bonding. Hahn et al. [5] provided the first experimental evidence that superlattice architectures can concurrently enhance hardness and fracture toughness, demonstrating a correlated peak in $H$ and $K_{IC}$ as a function of bilayer period in CrN/TiN superlattices. As emphasized in Ref. [1], only a limited number of experimental studies report such concurrent improvements, and the underlying mechanisms remain insufficiently validated. First-principles studies offer critical guidance, as they enable systematic screening of elastic anisotropy, shear-to-bulk modulus ratios ($G/B$), and Cauchy pressures ($P_{Cauchy}$), which correlate with the ability of a material to accommodate inelastic deformation ahead of a crack tip.

A comprehensive first-principles assessment of nitride/nitride superlattices identified a small subset of systems combining favourable elastic anisotropy, moderate $G/B$ ratios, and positive Cauchy pressures, leading to superior predicted fracture toughness [10]. Among the five highest-ranked systems based on calculated $K_{IC}$—$HfN/Ta_{0.75}N$, $TiN/WN_{0.75}$, $MoN_{0.5}/TaN$, $TiN/Ta_{0.75}N$, and $TiN/MoN_{0.5}$—three contain face-centred cubic (rocksalt, B1-type) TaN (δ-TaN) or defect-stabilized $Ta_{0.75}N$ layers, underscoring the decisive role of tantalum nitride sublayers in balancing stiffness and compliance at coherent interfaces. Notably, several of the remaining top-ranked systems, including TiN/WN [11], MoN/TaN [12], TiN/MoN [13], have already been experimentally validated, lending strong support to the underlying design principles and further highlighting TaN-based superlattices as particularly promising yet not fully explored candidates.

Importantly, TiN/TaN emerges from this analysis as a particularly attractive and historically relevant model system, combining a large elastic contrast with the ability to stabilize cubic $TaN_y$ through controlled substoichiometry and epitaxial constraint [10,14,15]. Beyond elastic considerations, tantalum nitride is distinguished by its pronounced structural metastability and exceptional defect tolerance, which fundamentally differentiate it from most chemically stable transition-metal nitrides. First-principles studies [16,17] have demonstrated that several Ta–N crystal structures—including hexagonal (ε-TaN), cubic rocksalt (δ-TaN), and alternative stacking variants—lie very close in energy of formation, explaining the strong sensitivity of experimentally observed phases to synthesis and growth conditions. Ab initio phase stability analyses [18] further reveal that stoichiometric δ-TaN is rarely the thermodynamic ground state, whereas Ta-deficient compositions ($Ta_{0.75–0.8}N$) are energetically favoured and characterized by a shallow Gibbs free energy landscape with respect to composition. This intrinsic tendency toward off-stoichiometry is accompanied by a strong preference for tantalum vacancies, rendering cubic TaN highly defect tolerant under non-equilibrium deposition conditions.

Early ion-assisted deposition studies demonstrated that intense nitrogen ion irradiation stabilizes metastable δ-$TaN_y$ at low temperatures through extreme non-equilibrium conditions, promoting nitrogen superstoichiometry, metal sublattice deficiency, and preferential ⟨100⟩-textured growth [19]. Consistent with this intrinsic metastability, epitaxial growth studies subsequently reported a high density of stacking faults and extended defects in cubic δ-TaN as a function of nitrogen activity and ion energy, even in nominally single-phase films [20]. The near-degeneracy of competing stacking sequences and the energetic favourability of metal vacancies imply low stacking-fault energies and a pronounced propensity for faulting and defect-mediated strain accommodation when cubic TaN is epitaxially constrained. Upon incorporation into superlattice architectures, these characteristics render TaN mechanically active rather than intrinsically brittle, enabling localized stress relaxation and interface-assisted deformation processes that are inaccessible in chemically more rigid nitrides. TiN/TaN superlattices therefore provide a unique platform in which metastability, defect tolerance, and elastic contrast can be deliberately exploited by compositional tuning to enhance fracture resistance without sacrificing hardness.

Beyond its favourable elastic metrics, TiN/TaN is distinguished by its propensity for stacking disorder, originating from the metastable rocksalt-to-hexagonal transformation tendency of TaN. Early experimental studies on TiN/TaN superlattices already revealed pronounced stacking irregularities and faulted sequences within the TaN layers, even though fracture toughness was not addressed explicitly at that time [15]. Subsequent high-resolution transmission electron microscopy studies demonstrated that these stacking faults can extend across interfaces and interact with the TiN layers, inducing localized lattice reorientation and defect-mediated plastic accommodation [21,22]. Such defect configurations are conceptually analogous to stacking-fault- and twin-assisted plasticity in metallic systems, where controlled faulting enables enhanced damage tolerance without sacrificing strength. While direct mechanical twinning in ceramic nitrides remains limited, stacking-fault-mediated deformation provides a viable pathway to locally relax stress concentrations and retard crack propagation.

Beyond structural design, deliberate interfacial doping with light non-metallic elements introduces an additional degree of freedom to tailor bonding strength, coherency strain, and local defect populations without altering the overall superlattice architecture. Atomically resolved investigations of multicomponent nitride coatings demonstrate distinct dopant distribution behaviours, with carbon and boron preferentially incorporating into nitride lattices, whereas silicon exhibits a strong tendency to segregate to interfaces and grain boundaries. Such interfacial segregation sharpens chemical gradients, enhances interface cohesion, and modifies local elastic mismatch, thereby directly influencing crack initiation and propagation behaviour [23].

Here, we investigate TiN/TaN superlattices with bilayer periods of approximately 6 nm and introduce atomic-scale interfacial doping with C, B, or Si to explore interface chemistry as an additional design parameter for tailoring mechanical performance. Combining X-ray diffraction, transmission electron microscopy, elastic recoil detection analysis, atom probe tomography, nanoindentation, micromechanical fracture testing, and first-principles calculations, we establish structure–property relationships linking interfacial chemistry to elastic response, defect evolution, and fracture resistance. By comparing the distinct effects of Si, C, and B, we demonstrate how interface chemistry selectively promotes either interface strengthening or defect-assisted energy dissipation, providing a design strategy for simultaneously enhancing hardness and fracture toughness in ceramic superlattices.

## 2 Experimental details

### 2.1 Coating deposition

TiN, TaN, and undoped as well as C, B, and Si-doped TiN/TaN superlattices were deposited by reactive unbalanced magnetron sputtering using an AJA Orion 5 system equipped with a 3-inch Ti and 2-inch Ta and dopant (C, B, Si) targets. The Ti target surface normal was inclined by 15° from the substrate-holder normal, whereas the Ta and dopant target surface normals were inclined by 12°. The target-to-substrate-holder distance was ~70 mm, with the target axes directed toward the central region of the substrate holder, while the substrate holder was continuously rotated at ~1 Hz throughout the deposition process. Substrates consisted of (001)-oriented MgO and Si platelets ($10 \times 10 \times 0.5$ mm$^3$), which were ultrasonically cleaned in acetone and ethanol before being mounted in the deposition chamber. Prior to deposition, the substrates were thermally cleaned at 500 °C for 20 min under vacuum (base pressure < 0.1 mPa) and subsequently plasma-etched in Ar using a high-energy DC glow discharge (−750 V substrate bias, total pressure 6 Pa) to remove residual surface contaminants and promote film adhesion.

Monolithic TiN and TaN films were initially deposited to calibrate deposition rates and optimize process parameters. Stabilization of the cubic δ-TaN phase required precise control of the nitrogen fraction in the reactive gas mixture, defined as $f_{N2} = F_{N2}/(F_{Ar} + F_{N2})$, where $F_{N2}$ and $F_{Ar}$ denote the $N_2$ and Ar flow rates, respectively. The nitrogen fraction ($f_{N2}$) was varied between 0.2 and 0.5 by adjusting $F_{N2}$ and $F_{Ar}$ between 2–5 sccm while maintaining a constant total flow of 10 sccm. Deposition pressure (0.4 Pa), substrate temperature (500 °C), and DC bias (−60 V) were kept constant. Based on these optimization experiments, the superlattice coatings were deposited at $f_{N2}$ = 0.3 ($F_{N2}$ = 3 sccm, $F_{Ar}$ = 7 sccm), corresponding to a nitrogen partial pressure of 0.14 Pa. Ti and Ta powers were 400 W (9.1 W/cm$^2$) and 200 W (10.2 W/cm$^2$), respectively. The dopant cathodes were operated at lower powers (B and Si: 50 W, 2.5 W/cm$^2$; C: 25 W, 1.3 W/cm$^2$) during interface modification.

Atomically thin layers of C, B, or Si were introduced specifically at the TiN/TaN interfaces using computer-controlled shutter sequences. For each interface, the respective 2-inch dopant cathode was opened for 1 s at low power, depositing an ultrathin dopant layer before the deposition of the adjacent TiN or TaN layer was resumed. The dopant cathodes were operated at 50 W (2.5 W/cm$^2$) for B and Si and 25 W (1.3 W/cm$^2$) for C. The B cathode showed a measurable C signal in energy-dispersive X-ray spectroscopy (EDS). The resulting dopant concentrations were not independently optimized but reflect the different deposition and incorporation rates of the respective dopant species under the identical, computer-controlled interface-decoration procedure. This approach was selected to introduce a reproducible, low-level chemical perturbation at each interface rather than a continuous doped layer. The superlattices were deposited with a nominal TiN:TaN thickness ratio of 1:1 and a bilayer period Λ of approximately 6 nm, resulting in total film thicknesses of ~2 μm.

### 2.2 Chemical and structural characterization

Film morphology and thickness were examined by scanning electron microscopy (SEM, FEI Quanta 250, 10 kV), and chemical composition was assessed using EDS (Philips XL30). Crystalline structures were characterized by X-ray diffraction (XRD) in Bragg–Brentano geometry using Cu Kα radiation (45 kV, 40 mA). Detailed microstructural analysis and interface quality were examined by high-resolution transmission electron microscopy (HRTEM) using a JEOL 2100F field-emission instrument equipped with an image-side spherical aberration corrector. Imaging was performed at 200 kV with a resolution of 1.2 Å, and the spherical aberration coefficient was tuned to near zero (Cs ≈ 0 μm). All high-resolution images were acquired under slight overfocus conditions. Together, these complementary techniques provided a comprehensive assessment of layer integrity, interface sharpness, and dopant confinement within the TiN/TaN-based superlattices.

The chemical composition of the superlattices was determined by ion beam analysis at the Tandem Laboratory of Uppsala University [24]. Depth profiling has been carried out using time-of-flight elastic recoil detection analysis (ERDA) and recoils were generated with a $^{127}I^{8+}$ ion beam at 36 MeV energy. The detection telescope consisted of thin carbon foils for measuring the time-of-flight [25] as well as a gas detection system [26]. Systematic uncertainties originate from the detection efficiency <1 for light recoils such as N and the specific energy loss of primary ions and recoil species [27]. Stoichiometric TiN [28] was used as reference sample and the measurement uncertainty was 3% relative of the deduced values. Assuming stoichiometric TiN/TaN and densities of 5.39 $g \cdot cm^{-3}$ for TiN [29] as well as 14.97 $g \cdot cm^{-3}$ for TaN [29], a depth of ~260 nm was probed. Average compositions were quantified from the surface-near region (~20–40 nm) of the depth profiles to minimize the influence of multiple scattering, caused by Ta, on the quantification [30]. This analysis region corresponds to the later stages of film growth, after stabilization of the deposition process, and therefore represents the steady-state coating composition.

The nanoscale composition of the superlattices close to the film/substrate interface was characterized by atom probe tomography (APT, Cameca LEAP 4000X HR). Field evaporation was assisted by thermal pulsing with a UV laser. 30 pJ laser pulse energy, 125 kHz laser pulse frequency, 60 K base temperature and 0.5% detection rate were employed. Reconstructions were created based on the shank angle protocol and the bilayer period $\Lambda$ of ~6 nm using AP Suite 6.1. Atom probe specimens were fabricated through focused ion beam (FIB) techniques in a dual-beam microscope (FEI Helios Nanolab 660) using Ga ions. Because APT specimens were extracted close to the film/substrate interface, the measured compositions locally reflect the initial stages of deposition, where transient oxygen incorporation from residual surface contamination, target conditioning, or chamber background may contribute more strongly than during steady-state growth. Therefore, differences in oxygen concentration between ERDA and APT should be interpreted as depth-dependent incorporation behaviour rather than as compositional inconsistencies between the techniques.

### 2.3 Mechanical properties

Hardness $H$ and indentation modulus $E$ were measured by nanoindentation (UMIS, Berkovich tip) under load-controlled conditions with maximum forces of 3–45 mN, with indentation depths below 10% of the coating thickness to minimize substrate influence [31]. The indenter was calibrated on fused silica ($E$ = 72.5 GPa) following Fischer-Cripps [32] and verified on additional reference samples as in Ref. [33]. H and E were extracted from load–displacement curves using the Oliver–Pharr method [34], with analysis conventions and data-quality checks following best practices summarized by Fischer-Cripps [35]. For each condition, n = 31 indents were performed; values are reported as mean ± one standard deviation.

Fracture toughness ($K_{IC}$) was determined by microcantilever bending tests performed in a SEM/FIB setup (FemtoTools FT-NMT04). Freestanding cantilevers ($w$:$b$:$l$ = 1:1:5 with width $w$, thickness $b$ and length $l$) were fabricated by precision FIB milling using a FEI Quanta 200 3D FIB. Here, $l$ refers to the distance between the pre-notch and the loading contact point; therefore, the actual cantilever length is slightly larger. The same FIB instrument was also used to remove the substrate material beneath the cantilevers. Material removal was carried out in successive milling steps with decreasing ion beam currents (5 nA, 3 nA, and 500 pA), while cantilever shaping was done with 1 nA for coarse milling and 500 pA for fine patterning to minimize ion-beam-induced damage. Pre-notches were introduced using a 50 pA ion beam to promote stable mode I crack propagation [36]. The notch depth $a_0$, and the dimensions $w$ and $b$ were measured post mortem from SEM micrographs of the fractured beams ($l$ was measured prior to the tests). For each condition, n = 5–6 cantilevers were tested; $K_{IC}$ values are reported as mean ± one standard deviation, and error bars in the figures reflect this scatter. Load–displacement curves were linear-elastic up to fracture (no detectable plasticity), consistent with brittle failure of the nitride layers. Our methodology and resulting $K_{IC}$ values are benchmarked against published TiN microcantilever data and fall within the reported ranges when using the same geometry factors [11,37]. All FIB processes were conducted at an acceleration voltage of 30 kV. In addition, the integrated energy dissipated during catastrophic fracture ($J_n$) was calculated from the load–displacement response as a comparative metric of fracture-energy dissipation and is not interpreted as an independent fracture toughness parameter.

This methodology allows direct correlation of mechanical performance with superlattice architecture and interfacial doping by providing intrinsic measurements of hardness, indentation modulus, and fracture toughness, while the integrated fracture energy serves as an additional descriptor of energy dissipation during failure.

### 2.4 Computational methods

Density Functional Theory (DFT) calculations were performed with the aid of the Vienna Ab-initio Simulation Package (VASP) [38,39] together with plane-wave projector augmented wave (PAW) pseudopotentials [40] and the Perdew-Burke-Ernzerhof generalized gradient approximation (GGA) [41]. The plane-wave cutoff energy of 600 eV and the reciprocal space sampling with Γ-centred Monkhorst-Pack meshes [42] ensured a total energy accuracy of at least $10^{-3}$ eV/at. Equilibrium lattice parameters of the cubic rocksalt (B1, $Fm\bar{3}m$) TiN and TaN were extracted from fitting the minimum of the energy–volume curve. Ta- and/or N-substoichiometric $Ta_xN$, $TaN_y$, as well as $Ta_xN_y$ structures were modelled in 64-atom supercells (≈0.83 $nm^3$), by removing a desired number of atoms in a quasi-random (SQS [43]) manner. TiN/TaN interfaces (with 1:1 TiN-to-TaN ratios, (001) orientation, and bilayer period of 1.6–1.9 nm) were modelled in 128-atom supercells, in which Ta or N vacancies were generated in the same way as for bulk systems. To study the effect of interface impurities (C, N, O, or Si), two impurity atoms (of the same type, yielding ≈ 1.6% impurity concentration) were placed at the Ta or N vacancy positions within the interface plane (with total of 8 metallic and 8 non-metallic fcc lattice sites). All vacancy-containing systems and superlattices—including those with impurities—were fully relaxed in terms of volume, cell shape and atomic positions. Relative chemical stability of relaxed systems was estimated by the formation energy,

$$E_f = \frac{E_{tot} - n_{Ti}\mu_{Ti} - n_{Ta}\mu_{Ta} - n_N\mu_N}{n_{Ti} + n_{Ta} + n_N}, \quad (1)$$

where $E_{tot}$ is total energy of the system, $n_{Ti}$ ($n_{Ta}$, $n_N$) is the number of Ti (Ta, N) atoms, and $\mu_{Ti}$ ($\mu_{Ta}$, $\mu_N$) the corresponding chemical potential, conventionally set to total energy per atom of the hcp-Ti (bcc-Ta, $N_2$ molecule). Mechanical properties were assessed based on elastic constants, $C_{ij}$, calculations (derived using the stress-strain approach [44,45]) and standard formulas for elastic moduli (bulk, shear, and elastic modulus, Young's modulus). Relative tendency for brittle/ductile behaviour was indicated by calculating the Cauchy pressure [46,47], defined as $C_{12} - C_{44}$. Mechanical strength of selected systems was further estimated by calculating the cleavage energy, $E_c$, and cleavage stress, $\sigma_c$, following Refs. [48,49] and considering cleavage along (001) planes. The Griffith's formula for fracture toughness, $K_{IC}$ [50],

$$K^{IC}_{[001]} = \sqrt{4E^c_{(001)}E_{[001]}}, \quad (2)$$

was employed, where $E^c_{(001)}$ is the cleavage energy and $E_{[001]}$ the directional elastic modulus. The latter was obtained from the calculated elastic constants using the standard relation for cubic crystals, $E_{[001]} = (C_{11}-C_{12})(C_{11}+2C_{12})/(C_{11}+C_{12})$.

# 3 Results

## 3.1 Composition and Structure

To establish suitable growth conditions for TiN/TaN SLs, we first investigated monolithic Ta–N films deposited at constant total pressure while systematically varying the nitrogen flow rate ratio $f_{N2}$. The Ta–N system exhibits pronounced sensitivity to nitrogen partial pressure, where small variations in $p_{N2}$ induce distinct phase fields, ranging from α-Ta at very low $p_{N2}$, via orthorhombic $Ta_4N$ and γ-$Ta_2N$ at intermediate values, to cubic δ-TaN or hexagonal ε-TaN under nitrogen-rich conditions [51,52,53].

X-ray diffraction patterns collected in Bragg–Brentano geometry reveal systematic changes in phase constitution, lattice parameter, and preferred orientation with increasing $f_{N2}$. In contrast to TiN, which retains a stable rocksalt structure over a wide nitrogen range, stabilization of cubic δ-TaN requires precise control of nitrogen supply. Since coherent TiN/TaN superlattice growth relies on crystallographic compatibility and lattice matching, identifying the narrow processing window for single-phase δ-TaN is essential.

Figure 1a summarizes the structural evolution of TaN with increasing $f_{N2}$. At $f_{N2}$ = 0.2, a mixed-phase microstructure comprising orthorhombic $Ta_4N$ and cubic δ-TaN is observed. Increasing $f_{N2}$ to 0.3 yields a single-phase δ-TaN film, confirming stabilization of the rocksalt structure under moderately nitrogen-rich conditions. A further increase to $f_{N2}$ = 0.5 induces a pronounced texture transition from (200) to (111), reflected by an increase in the intensity ratio $I_{111}/(I_{111} + I_{200})$ from ~0.3 to nearly unity, Figure 1b, indicating the development of a strongly ⟨111⟩-oriented microstructure. Concurrently, the lattice parameter expands from 4.37 Å to 4.40 Å, consistent with enhanced nitrogen incorporation.

Qualitative EDS analysis (Figure 1b) reveals an initially steep increase in nitrogen content when increasing $f_{N2}$ from 0.2 to 0.3, consistent with the transition from a mixed $Ta_4N$+TaN microstructure to single-phase TaN. At higher $f_{N2}$, the nitrogen content increases only mildly but monotonically, from about 42 to 46 at%. While the EDS data would suggest nitrogen-deficient $TaN_y$ compositions, the more accurate ERDA measurements (done on TiN/TaN superlattices) indicate the presence of superstoichiometric δ-TaN. Such compositions imply a high concentration of metal vacancies or Schottky-type defects, for example $Ta_{0.75}N$ and $Ta_{0.875}N_{0.875}$.

Density functional theory calculations (Table 1), in agreement with our previous work [10], show that increasing metal-vacancy concentrations reduce the ideal shear resistance of δ-TaN, reflecting a lowering of the intrinsic lattice stability. However, vacancy-rich rocksalt nitrides do not necessarily exhibit inferior fracture resistance, as Schottky defects can facilitate defect-mediated deformation mechanisms. In particular, our earlier study on Schottky-defected MoN [54] demonstrated that such defects promote the activation of partial dislocations under mechanical loading, resulting in enhanced experimentally measured fracture toughness.

Consequently, phase purity alone is insufficient to predict mechanical performance. Instead, an optimized balance between phase stability, vacancy population, texture development, and lattice coherency with TiN is required. Based on these combined criteria, $f_{N2}$ = 0.3 was selected as the optimal condition for the deposition of TiN/TaN superlattices. Figure 1a therefore presents also the XRD patterns of monolithically grown TiN and TaN films deposited under these conditions on MgO(001). Both coatings exhibit a pronounced 001 out-of-plane texture, with TiN growing almost exclusively in the 200 orientation and only very weak additional reflections at the 111 and 311 positions. TaN likewise shows a dominant 001 texture; the slightly more visible minor reflections arise from the logarithmic intensity scale used in Figure 1a and must not be overinterpreted.

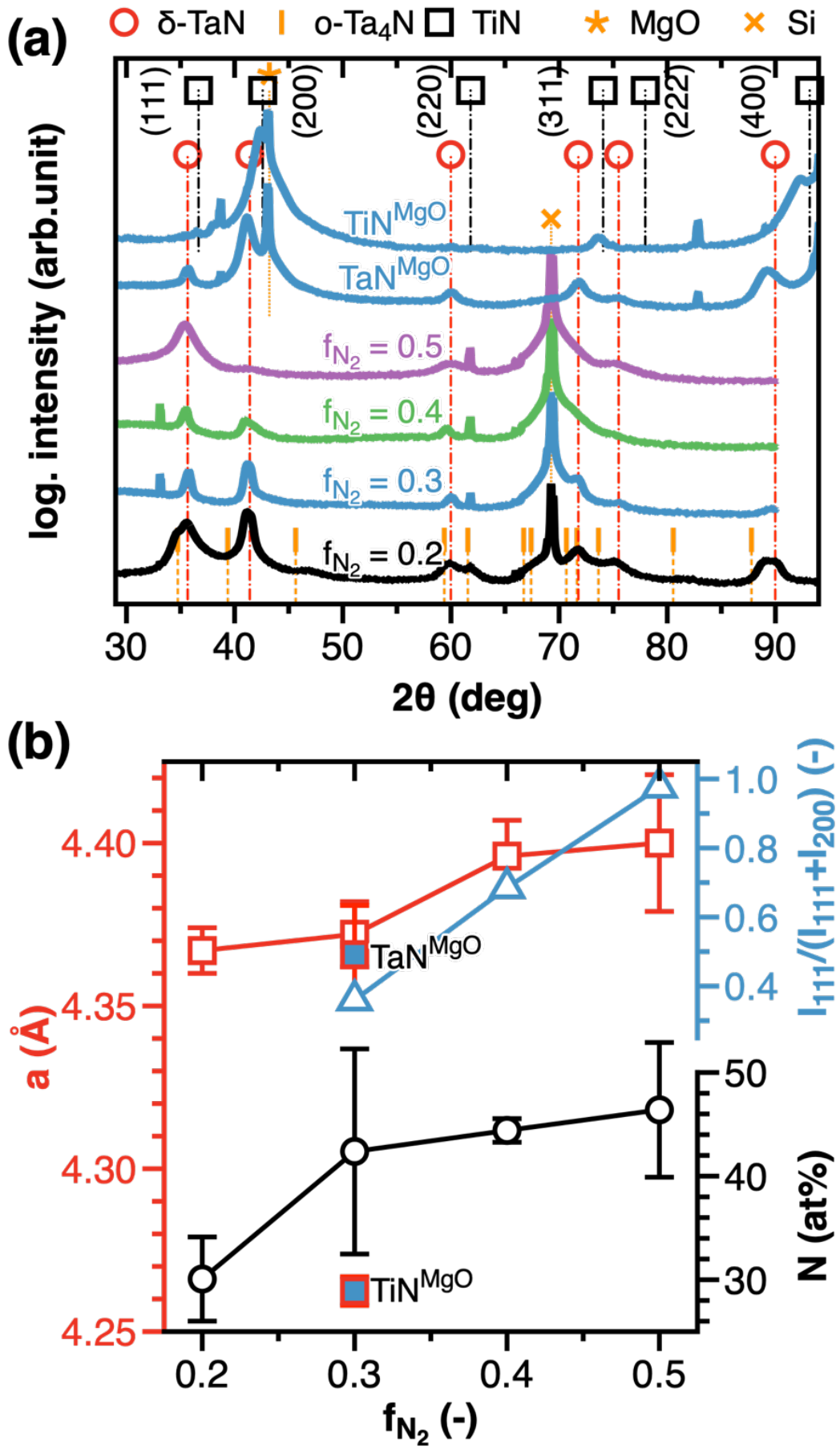

Figure 1 (a) XRD patterns of TaN coatings grown on Si(001) substrates at varying $N_2$ flow rate ratios ($f_{N2}$ = 0.2–0.5), together with TaN and TiN films deposited on MgO(001) at $f_{N2}$ = 0.3. Reference peak positions for orthorhombic o-$Ta_4N$ (ICDD #00-032-1282), cubic δ-TaN (ICDD #00-032-1283), and fcc TiN (ICDD #00-038-1420) are indicated, along with reflections from the Si and MgO substrates. (b) Lattice parameters derived from XRD (open red squares: films on Si; red squares filled with blue: films on MgO), XRD peak intensity ratio $I_{111}/(I_{111} + I_{200})$ (open blue triangles), and nitrogen content measured by EDS (open black circles) as a function of $f_{N2}$.

With $f_{N2}$ adjusted to stabilize stoichiometric TiN and cubic δ-TaN, TiN/TaN SLs were synthesized using a nominal bilayer period Λ of ~6 nm and a 1:1 thickness ratio. Interfacial doping with C, B, or Si was introduced as atomically thin layers at each TiN/TaN interface using computer-controlled shutter sequences, enabling precise localization of dopants at the heterointerfaces.

The formation of a well-defined superlattice architecture is evident from the XRD patterns shown in Figure 2a. All superlattices exhibit well-defined satellite reflections surrounding the fundamental (cumulative) rocksalt peaks, with satellites resolved up to the ±4th order for the 111, 200, and 400 reflections.

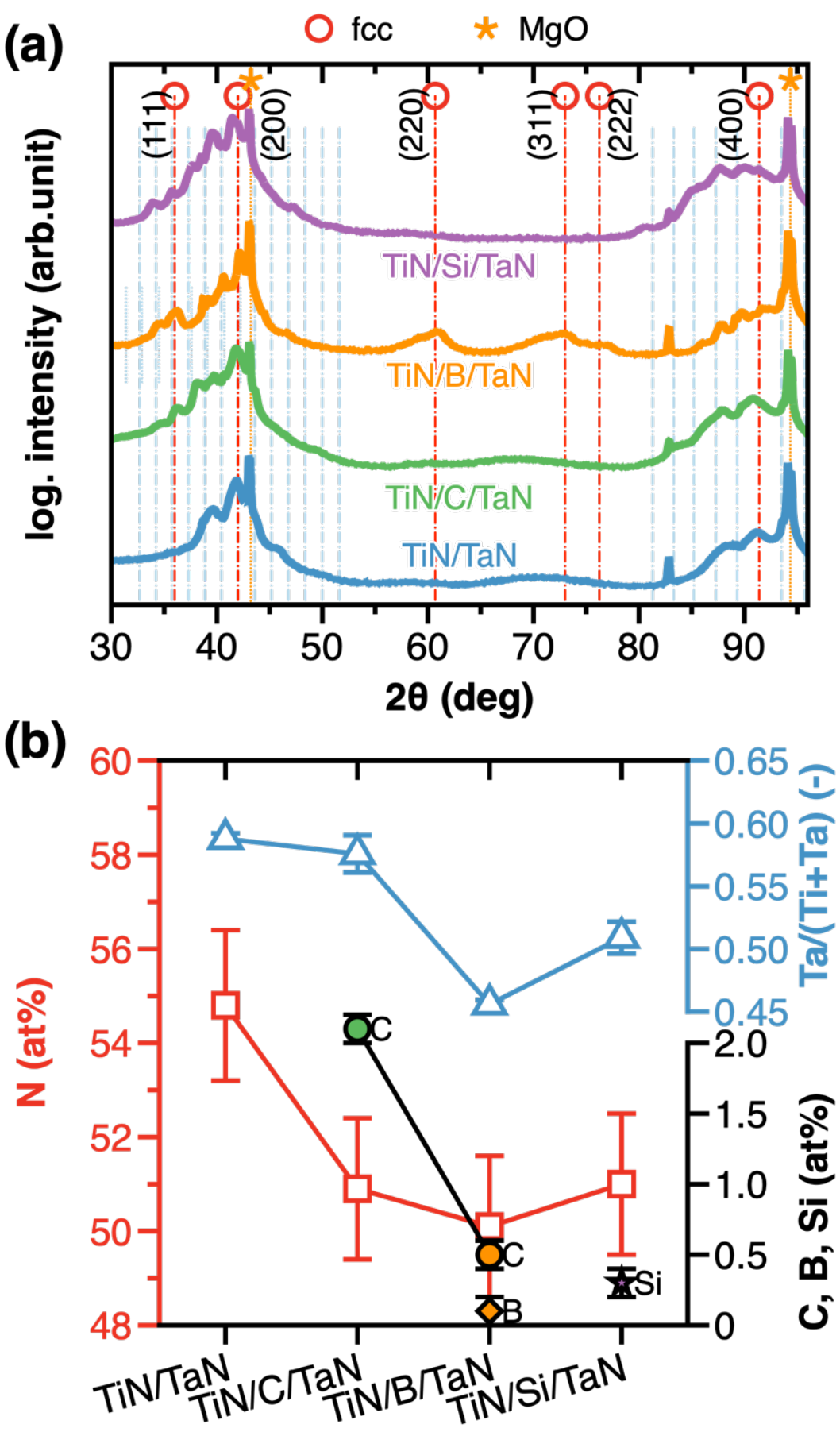

Figure 2: (a) XRD patterns of TiN/TaN, TiN/C/TaN, TiN/B/TaN, and TiN/Si/TaN superlattices grown on MgO(001). The position of the cumulative fcc reflection between TiN and TaN is indicated, together with superlattice satellite reflections up to $\pm 5^{th}$ order for the 111, 200, and 400 peaks, calculated for a bilayer period $\Lambda = 6$ nm. (b) Nitrogen content determined by ERDA (open red squares), Ta/(Ta + Ti) ratio measured by EDS (blue triangles), and C, B, or Si contents obtained by ERDA (black symbols with filling colour related to the diffractograms shown in (a)).

For the Si-doped superlattice, satellites are discernible even up to the ±5th order, most clearly for the 200 and 400 reflections. The superlattice satellite reflections exhibit a pronounced intensity asymmetry, with negative-order satellites being markedly stronger than their positive counterparts. This asymmetry is weak for the undoped TiN/TaN superlattice but becomes significantly more pronounced upon interfacial doping with C, B, or Si. At the same time, the regular angular spacing and sharpness of the satellite series confirm a well-defined bilayer period and coherent stacking of TiN and δ-TaN layers for all architectures (confirmed by TEM investigations, and exemplarily shown later for TiN/TaN and TiN/Si/TaN SLs). Given the larger lattice parameter and lower elastic stiffness of TaN compared to TiN, the enhanced asymmetry observed in the doped superlattices is consistent with increased strain and compositional modulation within the TaN-based layers. In combination with the measured nitrogen variation and dopant partitioning (Table 1 and Figure 2b), this behaviour indicates that dopants selectively amplify lattice distortion and elastic contrast in TaN while the TiN layers remain largely unaffected.

Table 1: Chemical compositions of the TiN/TaN-based coatings determined from ERDA depth profiling and EDS. Nitrogen, oxygen, argon, carbon, boron, and silicon concentrations were quantified by ERDA. The ERDA values represent the steady-state surface-near composition, sampled to depths of up to ~260 nm from the surface of the ~2 µm-thick films. The Ta/(Ta+Ti) ratio was determined by EDS, using both SEM top-view measurements and TEM cross-sectional line profiles. As the two measurement geometries yielded consistent values, the reported ratio represents their average, with the quoted uncertainty corresponding to the difference between the two measurements.

| film | EDS | ERDA | | | | | |
|---|---|---|---|---|---|---|---|
| | **Ta/(Ta+Ti)** | **N (at%)** | **O (at%)** | **Ar (at%)** | **C (at%)** | **B (at%)** | **Si (at%)** |
| TiN/TaN | 0.59±0.01 | 54.8±1.6 | 3.9±0.1 | 0.4±0.1 | - | - | - |
| TiN/C/TaN | 0.58±0.02 | 50.9±1.5 | 3.1±0.1 | 1.3±0.1 | 2.1±0.1 | - | - |
| TiN/B/TaN | 0.46±0.01 | 50.1±1.5 | 1.3±0.1 | 0.4±0.1 | 0.5±0.1 | 0.1±0.1 | - |
| TiN/Si/TaN | 0.51±0.01 | 51.0±1.5 | 0.8±0.1 | 1.0±0.1 | - | - | 0.3±0.1 |

The high satellite order and narrow peak widths indicate sharp compositional modulation and a high degree of long-range structural coherence, particularly for the doped SLs. Most SLs exhibit a dominant ⟨001⟩ orientation imposed by the MgO(001) substrate, while the TiN/B/TaN system additionally shows a weak ⟨111⟩ component and corresponding satellites, indicating a subtle modification of texture selection by interfacial B incorporation. The SL period Λ was determined from the satellite positions using the Eltoukhy–Greene relation,

$$\sin\theta_{\pm} = \sin\theta_B \pm m\lambda/(2\Lambda) \quad (3)$$

yielding Λ values of 5.2, 6.2, 6.2, and 6.6 nm for TiN/TaN, TiN/C/TaN, TiN/B/TaN, and TiN/Si/TaN, respectively, in close agreement with the nominal period derived from deposition rate calibrations and cross-sectional thickness measurements. The persistence of higher-order satellites across all samples confirms limited interlayer intermixing and low interface roughness, both prerequisites for pronounced superlattice strengthening effects.

Notably, the satellite reflections are significantly more pronounced for all doped SLs compared to the undoped TiN/TaN system, where they appear broadened and partially smeared. This enhanced satellite definition is attributed primarily to an increased chemical and elastic contrast between the TiN and TaN layers induced by interfacial doping, rather than to changes in interface roughness. Such enhanced layer distinctness is expected to influence coherency strain distribution and interface-mediated deformation processes discussed below. The preferential incorporation of C and B into TaN layers and the tendency of Si to diffuse into both adjacent layers (TiN and TaN), while also remaining at the interfaces sharpen the compositional modulation across the bilayers, thereby strengthening the superlattice diffraction response (this will be presented later when showing APT results).

The XRD patterns of the superlattices do not permit an unambiguous definition of a common lattice parameter for the individual TiN and TaN layers, given the finite bilayer period $\Lambda$ of ~6 nm. Instead, the dominant diffraction feature corresponds to a cumulative rocksalt reflection arising from the coherent superposition of TiN and TaN contributions. As shown in Figure 2a, the position of this cumulative peak lies between the expected fcc TiN and δ-TaN reflections, consistent with elastically constrained stacking of the two constituents within the superlattice architecture.

All SLs exhibit a pronounced ⟨001⟩ fibre texture imposed by the MgO(001) substrate, in agreement with the preferred orientation observed for monolithically grown TiN and TaN on MgO (Figure 1a). Notably, the formation of the TiN/TaN SL suppresses the minor secondary reflections (111, 220, 222, and 311) that are still detectable in monolithic TaN films grown on MgO, resulting in a highly texture-pure ⟨001⟩ response. An exception is observed for the B-doped TiN/B/TaN superlattice, where weak additional reflections at the 111, 220, 222, and 311 positions re-emerge together with their corresponding superlattice satellites (specifically for the 111 reflection). This behaviour indicates a locally disruptive effect of B incorporation on texture selection and stacking regularity, consistent with enhanced faulting or interface perturbation within the TaN layers.

Average compositions of the SLs as obtained by combining EDS and ERDA are summarized in Table 1. The undoped TiN/TaN superlattice is superstoichiometric with 55 at% N and exhibits a Ta/(Ta+Ti) ratio of 0.59. The N concentration is reduced to 51 at% for C- and Si-doped SLs, while the Ta/(Ta+Ti) ratios are 0.58 and 0.51, respectively. A further reduction to 50 at% and 0.46 is evident for the B-doped SL. The almost simultaneous change in Ta and nitrogen content for the SLs suggests that the TaN sublayers adopt a Ta-deficient composition of the form $Ta_xN$ (with $x < 1$).

The dopant concentrations determined by ERDA amount to approximately 2.1 at% C for TiN/C/TaN, 0.1 at% B for TiN/B/TaN (together with ~0.5 at% C, likely an impurity from the B target, which shows a detectable C signal in EDS), and 0.3 at% Si for TiN/Si/TaN. These results confirm that the introduced dopants remain at dilute levels, while still exerting a measurable influence on texture development and compositional balance. ERDA obtained oxygen contents are 3.9, 3.1, 1.3, and 0.8 at% for the TiN/TaN, TiN/C/TaN, TiN/B/TaN, and TiN/Si/TaN SLs, respectively. Collectively, the XRD and compositional data demonstrate that interfacial doping modifies stacking regularity and defect populations primarily through subtle perturbations of local chemistry and texture selection rather than through gross changes in lattice mismatch. The measured dopant concentrations are consistent with the intended atomic-scale interface doping, with no evidence of macroscopic segregation or bulk phase formation.

Figure 3 presents representative cross-sectional HRTEM images acquired at different magnifications, together with the corresponding fast Fourier transforms (FFTs), for undoped TiN/TaN and Si-doped TiN/Si/TaN superlattices grown on MgO(001). Both architectures exhibit a cubic rocksalt structure and maintain epitaxial registry across multiple bilayers, in agreement with the XRD results discussed above.

Overview HRTEM images (Figure 3a and c) resolve several superlattice periods and confirm a dense layered architecture with a bilayer thickness of ~6 nm. The undoped TiN/TaN superlattice shows highly uniform layer thickness and sharp interfaces with minimal contrast variations, indicating a high degree of structural coherence. In contrast, the Si-doped TiN/Si/TaN superlattice exhibits locally enhanced contrast within the TaN-based layers, suggesting increased lattice distortion and defect density associated with the interfacial Si incorporation.

Higher-magnification HRTEM images (Figure 3b and d) resolve the TiN/TaN interfaces at atomic resolution, capturing a single TaN layer coherently sandwiched between TiN layers in both architectures. FFTs extracted directly from the HRTEM images along the [100] zone axis show well-defined cubic reflections, confirming coherent epitaxial stacking across the superlattice interfaces. The FFTs of the Si-doped SL exhibit lower-intensity reflections together with an enhanced diffuse background compared with the undoped case. This reciprocal-space signature is consistent with elevated local lattice strain/disorder, which also reduces atom-by-atom image contrast in Fig. 3d despite identical acquisition conditions, and does not indicate a loss of long-range coherency. HRTEM measurements yield representative interplanar spacings of $d_{200} \approx 2.14$ Å for TiN and 2.18 Å for TaN. The measured superlattice periods are $\Lambda_{TEM} = 6.5$ nm for the Si-doped and 5.2 nm for the undoped SL, in good agreement with the corresponding periods from XRD satellite reflections ($\Lambda_{XRD} = 6.6$ nm and 5.2 nm, respectively).

Overall, the HRTEM analyses demonstrate that interfacial Si doping preserves coherent superlattice growth, while promoting localized structural disorder within the TaN-based layers. This combination of maintained epitaxy and enhanced defect-mediated strain accommodation produces a structurally heterogeneous yet coherent architecture, which provides a favourable foundation for the interface-assisted deformation and crack-tip shielding mechanisms discussed in the following sections.

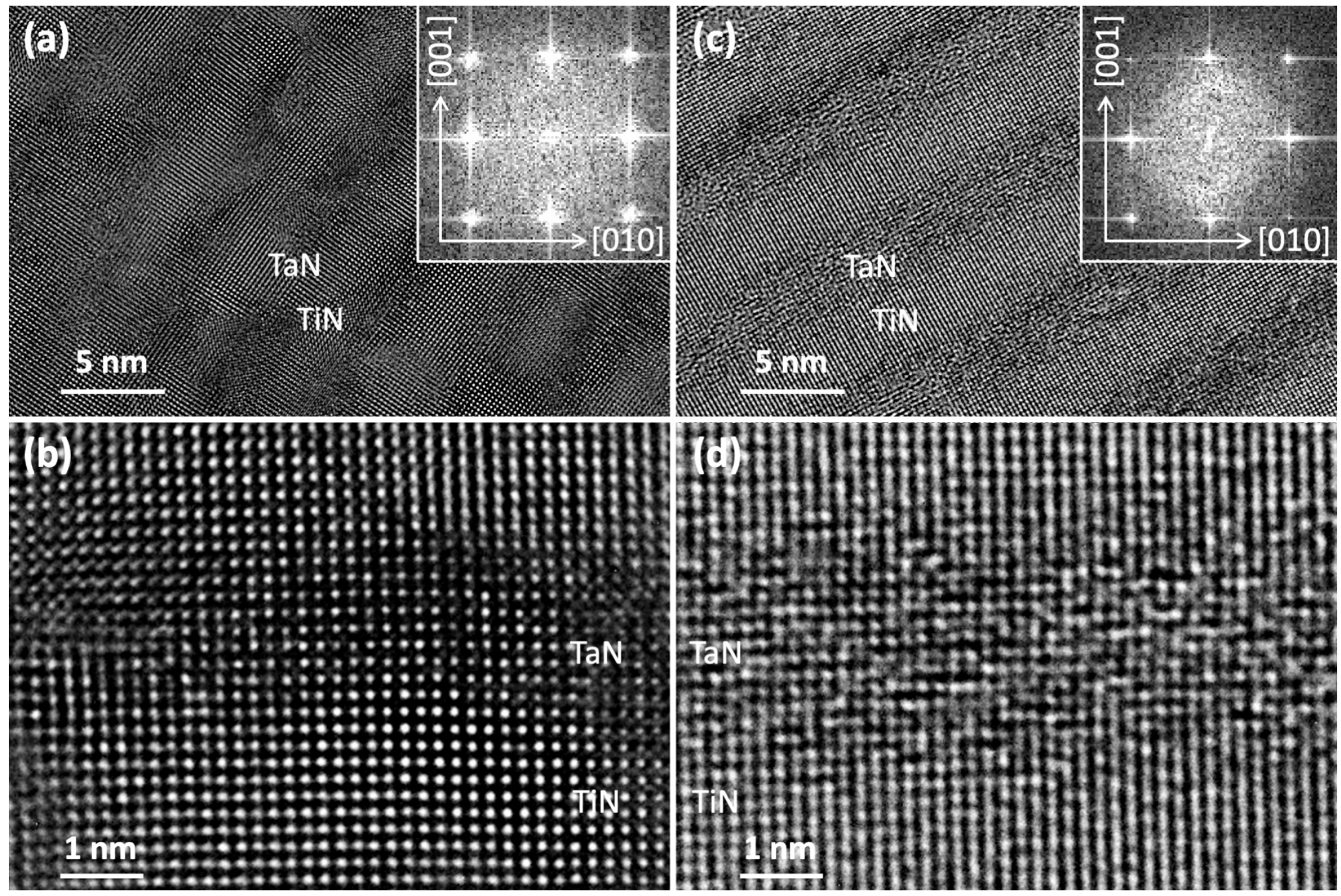

Figure 3: (double column) Cross-sectional HRTEM images of (a, b) TiN/TaN and (c, d) TiN/Si/TaN superlattices grown on MgO(001). Panels (a) and (c) show overview images comprising several bilayers, confirming the periodic superlattice architecture. Panels (b) and (d) resolve the TiN/TaN interface at atomic resolution, capturing a single TaN layer coherently sandwiched between TiN layers. Insets in (a) and (c) display the corresponding fast Fourier transforms (FFTs) taken along the [100] zone axis, confirming coherent cubic epitaxy across the superlattice. In addition, compared with the undoped case in (a), the FFT of the Si-doped SL in (c) exhibits broader, lower-intensity reflections together with an enhanced diffuse background, consistent with locally elevated lattice strain/disorder.

Figure 4 presents three-dimensional APT reconstructions (top row) together with one-dimensional composition depth profiles (bottom row) for the undoped and interfacially doped superlattices. In each panel, the black rectangle marks the 5 × 20 nm cylindrical region from which the one-dimensional profile was extracted. The TiN/TaN interfaces are identified by the cross-over of the Ti (blue) and Ta (cyan) signals. For the doped superlattices (Figs. 4b–d), 1 at% isoconcentration surfaces of C (green), B (orange), and Si (purple) highlight the corresponding dopant-enriched regions. Mass spectra are provided in Figures S1–S4 in the Supplementary Materials.

The APT data confirm the periodic TiN/TaN architecture in all investigated coatings, while revealing dopant-dependent intermixing and partitioning. In the undoped and C-doped superlattices, the Ti and Ta profiles alternate with relatively sharp transitions and limited intermixing. In contrast, the B- and Si-doped superlattices exhibit broader overlap of the Ti and Ta profiles, indicating increased intermixing across the interfaces. The dopant distributions follow distinct partitioning behaviour: C and B are detected predominantly within the TaN-based layers, whereas Si is enriched at the interfaces while also extending into both adjacent TiN and TaN layers. The maximum local dopant concentrations reach ~2 at% C in TiN/C/TaN, ~7 at% B in TiN/B/TaN, together with ~2 at% C, and ~2 at% Si in TiN/Si/TaN. The minor C signal detected in the B-doped coating is attributed to a small impurity contribution from the B source. These observations confirm that the nominally identical interface-decoration procedure produces distinct dopant distributions depending on the chemical nature of the dopant.

For Si, peak overlap with N occurs at 14 Da ($N^+$, $Si^{2+}$) and 28 Da ($N_2^+$, $Si^+$). Given the low average Si concentration of 0.3 at% (Table 1) determined by ERDA, the Si contribution to these mass peaks is difficult to separate reliably from N. The Si signal in the APT profiles was therefore primarily assigned using the 9.3, 22, and 44 Da peaks corresponding to $Si^{3+}$, $SiO_2^+$, and $SiO^+$, respectively. The resulting Si concentration in the one-dimensional profiles should consequently be regarded as a lower estimate of the local Si content.

A pronounced oxygen depth dependence is also observed in the APT datasets. All superlattices exhibit significant O incorporation in the film/substrate interfacial region analysed by APT, with local concentrations ranging from ~12 to ~32 at%. In the undoped and C-doped superlattices, for example, the O concentration decreases by approximately 10 at% over the analysed region, from about 30 at% to ~20 at% within ~20–40 nm. This behaviour is consistent with oxygen gettering by the Ti and Ta targets during the early stages of deposition. The detection of TiO and TaO molecular ions in the APT mass spectra (Figs. S1–S4) further supports this interpretation. As deposition proceeds, the gettering effect diminishes as residual oxygen and/or water vapour are progressively depleted from the chamber. Consequently, the surface-near O concentrations determined by ERDA (down to ~260 nm from the surface) after deposition of the ~2 μm-thick films are substantially lower, ranging from 3.9, 3.1, 1.3, and 0.8 at% for the undoped, C-, B-, and Si-doped superlattices, respectively. Oxygen gradients were also observed by EDS line scans across TEM lamellae (not shown). Interestingly, both APT and ERDA exhibit the same decreasing O-content trend across the deposition sequence, from the undoped to the C-, B-, and Si-doped coatings. This systematic evolution is consistent with progressive conditioning of the deposition system and target surfaces during sequential deposition, which increasingly reduces the residual oxygen and/or water-vapour background available for gettering. Importantly, despite the initially elevated O incorporation, the periodic Ti/Ta alternation and coherent interfaces are preserved, as demonstrated by the APT and HRTEM analyses (Fig. 3). From a crystallographic perspective, the close lattice-parameter match between TiN and δ-TaN supports coherent stacking over the selected bilayer period of approximately 6 nm. At this length scale, coherency is largely maintained, while localized strain accommodation and defect formation—particularly within the TaN layers—may already contribute to partial misfit relaxation, as suggested by the TEM analyses.

Argon concentrations of ~1–5 at% are detected by APT and are most likely associated with energetic Ar neutrals reflected from the Ta target [55], whereas ERDA yields substantially lower concentrations of ≤1.3 at%. TiO and TaO molecular ions are detected in the APT mass spectra of all samples, further supporting the proposed oxygen-gettering mechanism. Importantly, neither the oxygen nor argon incorporation disrupts the superlattice architecture: the periodic Ti/Ta alternation observed by APT, the sharp interfaces resolved by HRTEM, and the well-defined XRD satellite reflections demonstrate that the fundamental coherent TiN/TaN structure is preserved.

Overall, APT and ERDA establish a consistent picture of depth-dependent impurity incorporation and dopant-specific interface chemistry. C and B preferentially partition into the TaN-based layers, whereas Si remains enriched at the interfaces while also incorporating into both adjacent layers. These distinct chemical distributions provide the structural basis for interpreting the different elastic and fracture responses discussed below.

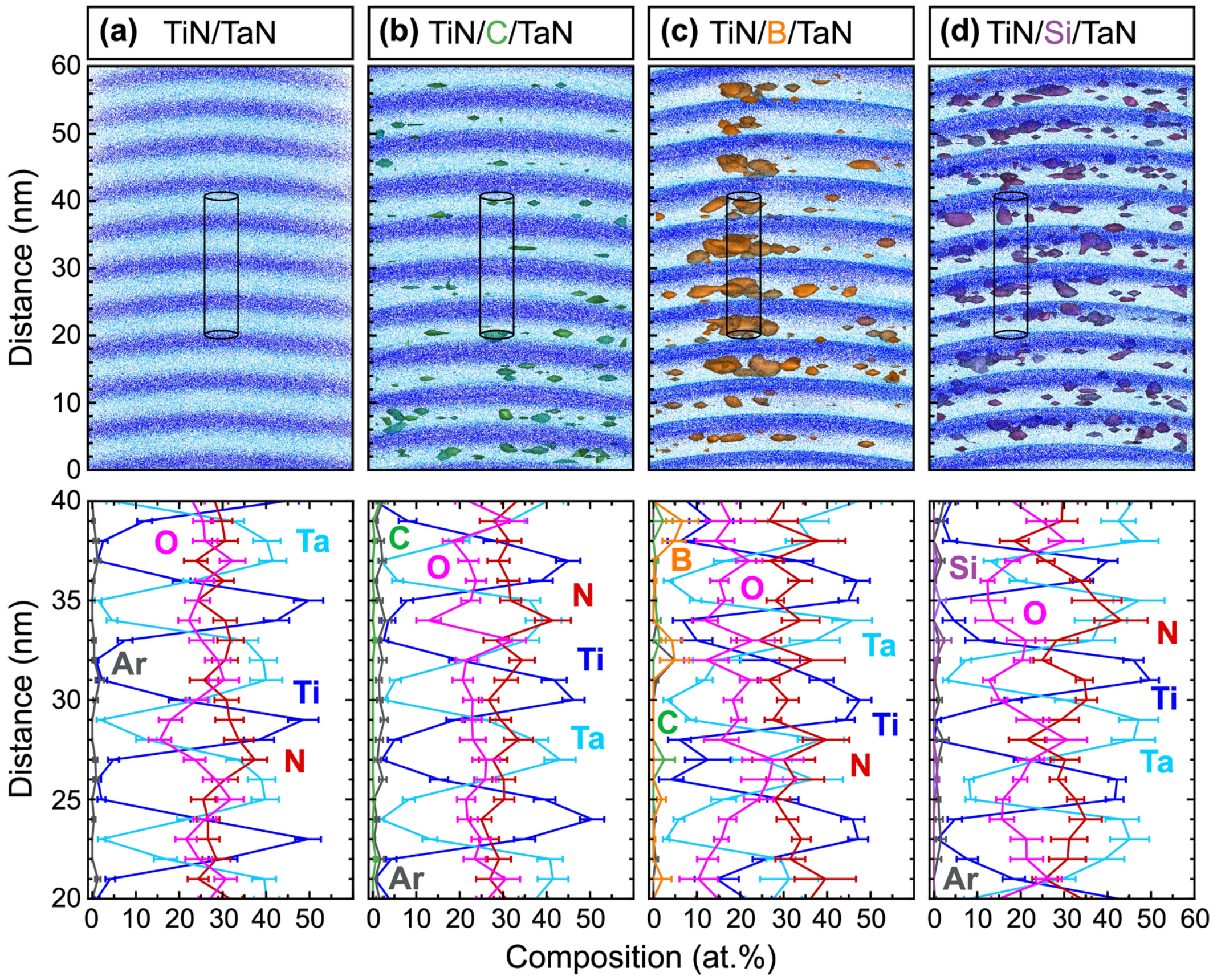

Figure 4: (double column) APT of (a) TiN/TaN, (b) TiN/C/TaN, (c) TiN/B/TaN, and (d) TiN/Si/TaN superlattices grown on MgO(001). Top row: 3D reconstructions showing Ti (blue) and Ta (cyan). In (b–d), dopant distributions are additionally highlighted as 1 at% isoconcentration surfaces: C (green), B (orange), and Si (purple). The black rectangles in (a–d) mark the 5 × 20 nm cylindrical regions used for the 1D analysis. Bottom row: one-dimensional composition depth profiles extracted along the cylinder axis. The TiN/TaN interfaces are identified by the cross-over of the Ti (blue) and Ta (cyan) signals. N, O, and Ar are shown in red, orange, and grey, respectively.

### 3.2 Indentation hardness and modulus

Because nanoindentation probes the surface-near region, with the indentation depths used here limited to ≤200 nm, the measured hardness and indentation modulus are most appropriately discussed in relation to the surface-near compositions determined by ERDA. ERDA probes to depths of up to ~260 nm and therefore provides a chemically relevant reference for the indentation measurements, in contrast to the substantially higher oxygen concentrations detected by APT near the film/substrate interface. Nanoindentation measurements reveal pronounced differences in hardness between monolithic coatings and TiN/TaN-based superlattices, as summarized in Figure 5a. Monolithic TiN exhibits a hardness of ~28.5 GPa, while TaN reaches about 30.6 GPa under the selected deposition conditions. The undoped TiN/TaN superlattice attains a higher hardness of 34.4 GPa, exceeding both monolithic reference coatings and demonstrating a clear superlattice strengthening effect arising from the periodic modulation of elastic and crystallographic properties.

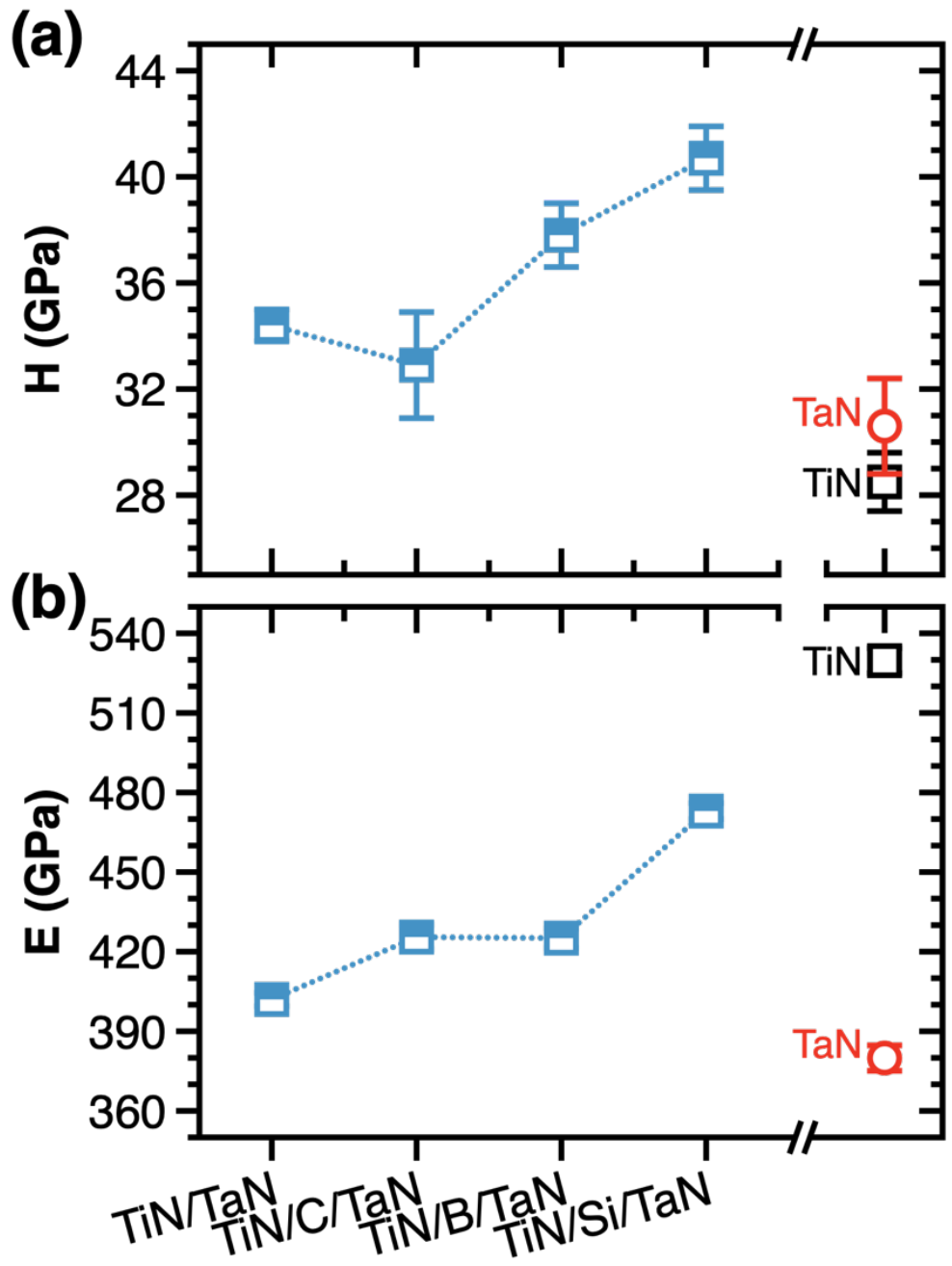

Figure 5: (a) Indentation hardness $H$ and (b) indentation modulus $E$ of TiN/TaN, TiN/C/TaN, TiN/B/TaN, and TiN/Si/TaN superlattices, with data of monolithically grown TiN and TaN included for comparison. Values represent the mean ± one standard deviation from n = 31 indents per condition, with error bars smaller than the symbol size not visible. All coatings were deposited on MgO(001) substrates.

Interfacial doping modifies the hardness in a pronounced, dopant-specific manner. Carbon-doped superlattices show hardness values comparable to the undoped architecture (32.9 GPa), indicating that C incorporation has only a minor influence on the dominant strengthening mechanisms. Boron doping leads to a moderate hardness increase to 37.8 GPa, whereas silicon doping produces the most pronounced effect, raising the hardness to approximately 40.7 GPa. This corresponds to an increase of ~20% relative to the undoped TiN/TaN superlattice.

The observed hardness trends correlate with the structural characteristics discussed in Section 3.1. XRD patterns show more pronounced superlattice satellite reflections for all doped architectures compared to the undoped TiN/TaN, indicating enhanced compositional modulation upon interfacial doping. Among the doped systems, the Si-doped SL exhibits the clearest higher-order satellite reflections, including weak fifth-order peaks, suggesting the highest degree of long-range periodicity. HRTEM directly confirms well-defined and continuous layer sequences for the undoped TiN/TaN and the Si-doped TiN/TaN superlattice, supporting the interpretation that Si doping most effectively stabilizes the periodic architecture. These observations suggest that the superior hardness of the Si-doped system originates from an optimized combination of periodicity, elastic contrast, and defects, rather than from a fundamental change of the TiN/TaN superlattice concept.

Figure 5b summarizes the indentation modulus values. Monolithic TiN exhibits an indentation modulus of approximately 530 GPa, whereas TaN shows a substantially lower value of ~380 GPa, in reasonable agreement with previous reports for sputtered TiN and TaN coatings [11,52]. The undoped TiN/TaN superlattice displays an effective indentation modulus close to that of TaN, reflecting the dominant contribution of the more compliant TaN layers to the overall elastic response.

Interfacial doping leads to a systematic increase in indentation modulus. Carbon- and B-doped superlattices both reach modulus values of approximately 425 GPa, while the Si-doped architecture exhibits a markedly higher modulus of ~472 GPa. This monotonic increase in modulus mirrors the evolution in hardness and indicates that interfacial dopants enhance the elastic stiffness of the superlattice without compromising structural integrity.

Density functional theory (DFT) calculations provide insight into the elastic response of TiN, TaN, and related coherent nitride superlattices (Table 2). The films are highly (001)-oriented, making directional elastic moduli $E_{[001]}$ particularly relevant. For TiN, the calculated $E_{[001]}$ of 524 GPa agrees well with the experimental (001)-oriented value of 530 GPa, validating the computational approach. In contrast, stoichiometric

TaN shows a much higher $E_{[001]}$ of 668 GPa, significantly exceeding the experimental 380 GPa. Vacancy-stabilized $Ta_xN_y$ models, however, reproduce the experimental modulus remarkably well: $Ta_{0.75}N$ yields 334 GPa, and $Ta_{0.875}N_{0.875}$ reaches 387 GPa, essentially matching the measured value. This agreement indicates that the sputtered TaN layers are intrinsically defected, likely via Schottky-type vacancies, which reduce the effective stiffness while preserving coherency with TiN. These observations underscore the importance of accounting for defect-stabilized TaN in understanding the mechanical behaviour and fracture resistance of TiN/TaN superlattices [10].

Table 2: Selected structural, elastic, and fracture-related properties calculated by density functional theory (DFT) for cubic (NaCl-type, B1) TiN, TaN, and selected nitride superlattices. For the superlattices, calculations correspond to coherent architectures with bilayer periods of 1.6–1.9 nm. Listed are the formation energy $E_f$ (eV/atom); lattice parameter mismatch $\Delta a$ (Å) and shear modulus mismatch $\Delta G$ (GPa) between the constituent layers; single-crystal elastic constants $C_{11}$, $C_{12}$, and $C_{44}$ (GPa); polycrystalline bulk ($B$), shear ($G$), and elastic modulus ($E$) (GPa); the $B/G$ ratio and Cauchy pressure ($C_{12} - C_{44}$, GPa) as indicators of bonding character; the directional elastic modulus $E_{[001]}$ (GPa); cleavage energy $E^c_{(001)}$ (J/m$^2$) and ideal cleavage stress $\sigma^c_{(001)}$ (GPa); and the orientation-dependent theoretical fracture toughness $K^{IC}_{[001]}$ (MPa√m) derived with Eq. 2. Off-stoichiometry in $Ta_xN_y$ is realized via vacancies. Part of the data comes from our previous work [10].

| material | $E_f$ | $\Delta a$ | $\Delta G$ | $C_{11}$ | $C_{12}$ | $C_{44}$ | $B$ | $G$ | $E$ | $G/B$ | $C_{12}$–$C_{44}$ | $E_{[001]}$ | $E^c_{(001)}$ | $\sigma^c_{(001)}$ | $K^{IC}_{[001]}$ |
|---|---|---|---|---|---|---|---|---|---|---|---|---|---|---|---|
| TiN | -1.77 | | | 573 | 132 | 162 | 279 | 183 | 451 | 0.66 | -30 | 524 | 3.2 | 31.3 | 2.6 |
| TaN | -0.91 | | | 717 | 146 | 69 | 340 | 127 | 338 | 0.38 | 77 | 668 | 3.1 | 27.6 | 2.9 |
| $Ta_{0.75}N$ | -1.03 | | | 417 | 154 | 91 | 242 | 105 | 276 | 0.42 | 84 | 334 | 3.9 | 34.6 | 2.3 |
| $TaN_{0.75}$ | -0.96 | | | 512 | 192 | 108 | 299 | 127 | 333 | 0.41 | 71 | 407 | 4.2 | 32.9 | 2.6 |
| $Ta_{0.875}N_{0.875}$ | -0.97 | | | 500 | 199 | 104 | 299 | 121 | 319 | 0.40 | 95 | 387 | 3.7 | 21.9 | 2.4 |
| $HfN/Ta_{0.75}N$ | -1.60 | 0.24 | 50 | 614 | 125 | 122 | 272 | 162 | 406 | 0.60 | 3 | 572 | 3.6 | 32.1 | 2.9 |
| $TiN/WN_{0.75}$ | -1.20 | 0.03 | 58 | 596 | 139 | 78 | 304 | 133 | 347 | 0.44 | 61 | 543 | 3.9 | 31.5 | 2.9 |
| $MoN_{0.5}/TaN$ | -0.72 | 0.21 | 4 | 609 | 161 | 62 | 302 | 106 | 285 | 0.35 | 99 | 542 | 4.0 | 32.0 | 3.0 |
| $TiN/Ta_{0.75}N$ | -1.55 | 0.05 | 78 | 614 | 132 | 144 | 277 | 173 | 431 | 0.62 | -12 | 567 | 3.5 | 31.6 | 2.8 |
| $TiN/MoN_{0.5}$ | -1.28 | 0.04 | 60 | 576 | 133 | 70 | 285 | 124 | 326 | 0.44 | 63 | 526 | 3.8 | 30.2 | 2.8 |

For TiN/TaN-based SLs, the experimentally observed increase in indentation modulus from 402 GPa for the undoped architecture to 425 GPa for the C- and B-doped systems and to 472 GPa for the Si-doped superlattice aligns with the DFT-predicted sensitivity of elastic properties to lattice and shear modulus mismatch between constituent layers (Table 2). These results indicate that interfacial doping provides an effective means to tune elastic contrast and coherency strain, thereby enhancing both hardness and stiffness while preserving the fundamental TiN/TaN superlattice architecture.

### 3.3 Fracture toughness evaluation

Figure 6a illustrates the experimental configuration for in-situ microcantilever bending, with a wedge indenter aligned above a free-standing cantilever prior to loading. A key aspect of the specimen design is the FIB-milled pre-notch, which is labelled with its depth $a_0$ in the cross-section shown in Figure 6b. The notch is defined by sacrificial material bridges on either side, which fracture during the initial stages of loading and ensure the formation of an atomically sharp crack tip. The notch depth $a_0$, together with the cantilever width $w$ and thickness $b$, was determined directly from SEM images and used for calculating the effective fracture surface area.

Figure 6c–h present post-mortem SEM images of the fracture surfaces of monolithic TiN and TaN films, as well as undoped and doped TiN/TaN SLs. In all cases, crack initiation occurs at the pre-notch, confirming controlled and reproducible fracture conditions. The fracture surfaces are smooth and featureless at the SEM scale, consistent with predominantly brittle fracture behaviour. While minor variations in surface contrast are observed, no systematic differences in fracture morphology can be resolved between the different architectures based solely on post-mortem SEM inspection.

Importantly, the smooth fracture surfaces validate the use of the projected fracture area in the fracture mechanics analysis. Any deviation between the projected and true fracture surface area remains small relative to the experimental scatter, ensuring that the observed differences in fracture toughness—ranging from about 2.2 MPa√m for TiN to nearly 4.0 MPa√m for B-doped TiN/TaN SLs (see below)—originate from intrinsic differences in crack-propagation resistance rather than from specimen geometry or surface roughness.

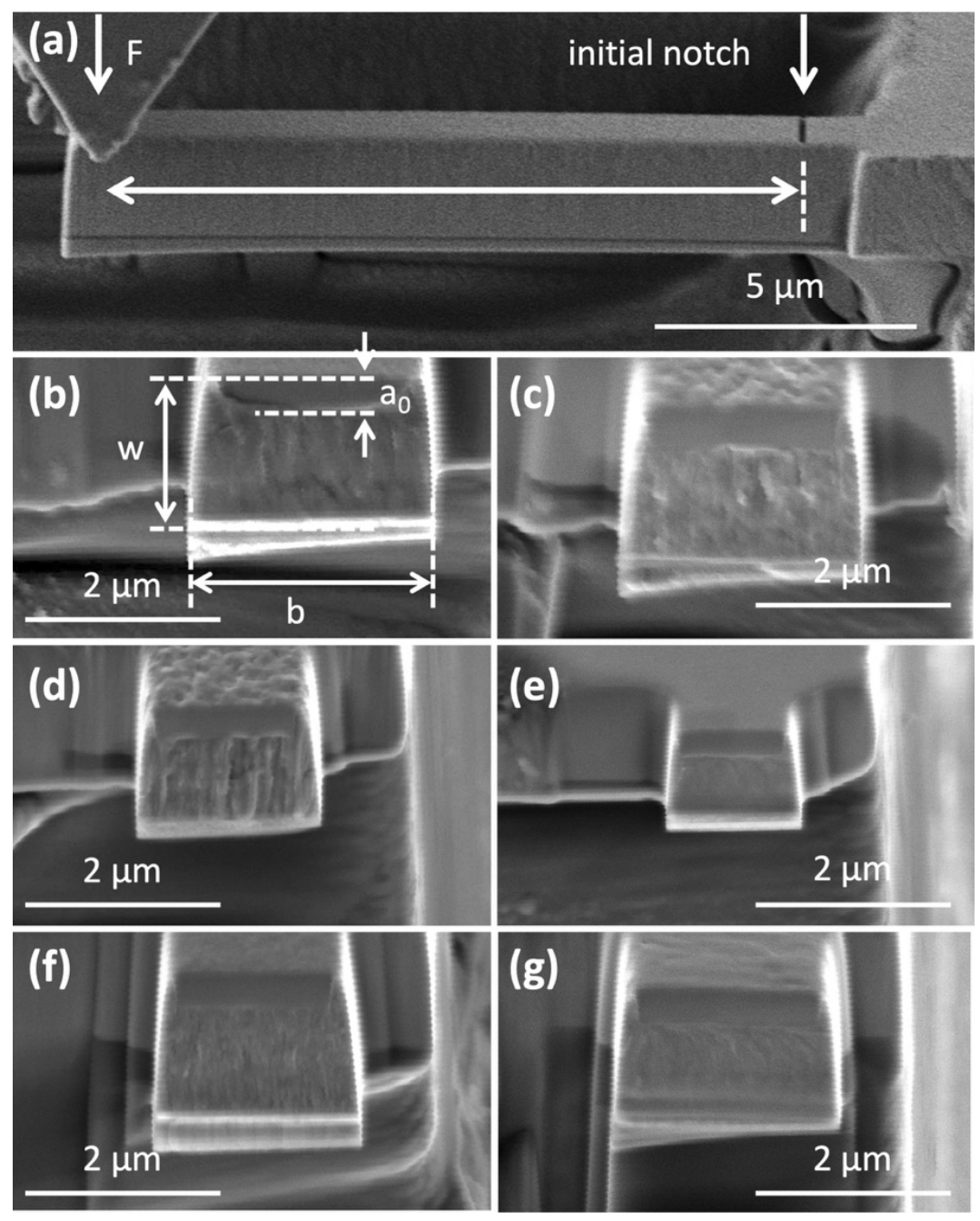

Figure 6: (a) SEM overview of a free-standing microcantilever fabricated by FIB milling and positioned for in-situ bending using a wedge indenter. (b) High-magnification SEM image of the FIB-milled pre-notch, showing the sacrificial material bridges on either side of the notch, which ensure the formation of an atomically sharp crack upon loading; the notch depth $a_0$ and the cantilever width $w$ and thickness $b$ used for fracture mechanics analysis are indicated. (c–h) Post-mortem SEM images of fracture surfaces after bending of monolithic TiN (c), monolithic TaN (d), and SLs of TiN/TaN (e), TiN/C/TaN (f), TiN/B/TaN (g), and TiN/Si/TaN (h). In all cases, fracture initiates at the pre-notch and proceeds in a predominantly brittle manner, resulting in smooth fracture surfaces.

Representative load–deflection curves (Figure 7) display linear elastic behaviour up to fracture for all investigated coatings, confirming the absence of measurable plastic deformation prior to crack propagation. Fracture toughness ($K_{IC}$) was therefore determined from the maximum load at fracture $F_m$ using established beam-bending relations for rectangular microcantilevers [56,57]:

$$K_{IC} = f(a_0/w)\frac{F_m l}{b w^{1.5}}, \quad (4)$$

with the geometry factor

$$f(a_0/\mathrm{w}) = 1.46 + 24.36(a_0/\mathrm{w}) - 47.21(a_0/\mathrm{w})^2 + 75.18(a_0/\mathrm{w})^3 \quad (5)$$

where $l$ is the distance from the notch to the point of force application, $w$ is the cantilever width, and $b$ is the cantilever thickness.

In addition to $K_{IC}$, the integrated fracture energy $J_n$ was calculated from the area under the load–deflection curve, representing the total elastic strain energy absorbed prior to fracture, normalized by the projected fracture surface area beneath the notch [58]:

$$J_n = \frac{\int F dx}{b \cdot (w - a_0)}, \tag{6}$$

where, $\int F dx$ is the total strain energy, and $b \cdot (w - a_0)$ is the projected fracture surface area under the notch.

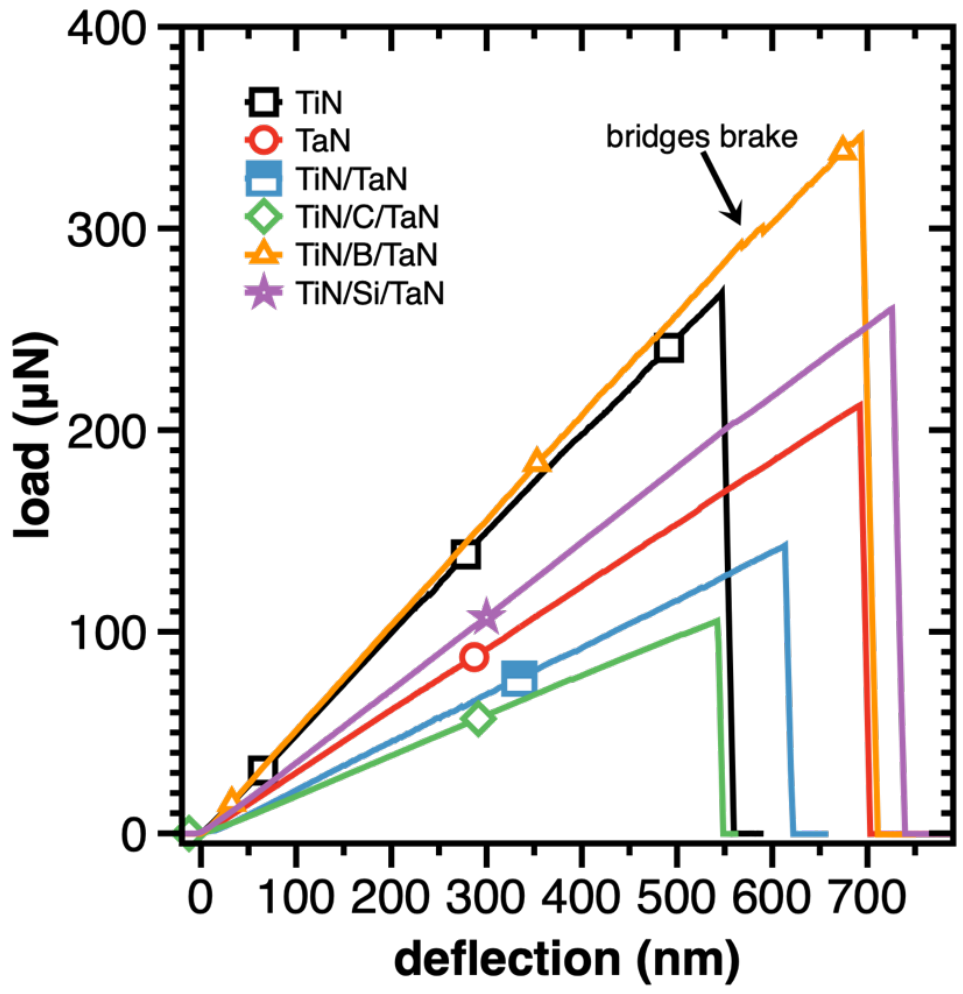

Figure 7: Representative load–deflection curves obtained from microcantilever bending tests of monolithic TiN and TaN, and TiN/TaN-based superlattices, showing linear elastic response up to catastrophic failure for all investigated coatings. All curves are linear-elastic up to fracture with no detectable plasticity.

Figure 8a and b summarize $K_{IC}$ and $J_n$ for monolithic coatings and TiN/TaN-based superlattices, respectively. Monolithic TiN and TaN exhibit fracture toughness values of 2.2 and 2.4 MPa√m, respectively, reflecting the intrinsically brittle nature of dense transition metal nitrides. The undoped TiN/TaN superlattice reaches a higher $K_{IC}$ of 2.8 MPa√m, confirming the beneficial effect of nanoscale layering on crack resistance.

Interfacial doping leads to a pronounced, dopant-specific enhancement of fracture toughness. Carbon-doped superlattices reach a $K_{IC}$ of 3.5 MPa√m, while boron doping yields the highest value of approximately 4.0 MPa√m. Silicon-doped superlattices also show a substantial increase in toughness, reaching 3.4 MPa√m. The same ranking is reflected in the measured fracture-energy dissipation: the integrated energy $J_n$ increases systematically upon doping, with the B-doped architecture exhibiting the highest energy absorption prior to catastrophic fracture. Although $J_n$ is not interpreted here as an independent fracture toughness parameter, its evolution is fully consistent with the corresponding increases in $K_{IC}$, indicating enhanced energy dissipation during fracture.

The fracture resistance of the investigated coatings reflects a non-monotonic interplay between bonding character, elastic compliance, and defect concentration. First-principles calculations identify TiN as intrinsically brittle, with a theoretical fracture toughness of ~2.5 MPa√m, a relatively high *G*/*B* ratio (~0.66), and a strongly negative Cauchy pressure (Table 2), consistent with predominantly directional covalent bonding and limited shear accommodation. In contrast, TaN exhibits a moderately higher calculated fracture toughness of ~2.8 MPa√m together with a lower G/B ratio and a positive Cauchy pressure, reflecting an increased metallic bonding contribution and greater elastic compliance. While *G*/*B* alone is not expected to quantitatively predict fracture toughness for these closely related nitride systems, the combined trends in *G*/*B*, Cauchy pressure, and elastic constants consistently indicate an enhanced ability of TaN to accommodate local shear deformation.

Moderate deviations from stoichiometry in $Ta_xN_y$ alter the elastic response but do not lead to a monotonic increase in fracture toughness. While Ta vacancy–stabilized compositions show reduced shear moduli and strongly positive Cauchy pressures, their calculated $K_{IC}$ values remain comparable to, or slightly below, that of stoichiometric TaN, indicating that excessive vacancy concentrations may compromise shear resistance despite enhanced compliance. These results suggest that stoichiometric or near-stoichiometric TaN occupies a favourable region in the elastic–fracture design space, combining sufficient stiffness with improved damage tolerance relative to TiN, whereas vacancy stabilization primarily modifies elastic contrast rather than intrinsically enhancing fracture resistance.

The systematic increase in $K_{IC}$ and $J_n$ upon interfacial doping supports this interpretation. Dopant-induced modifications of the local elastic response, combined with changes in bonding character at the interfaces, promote a more homogeneous stress distribution and delay unstable crack propagation. Boron, which predominantly incorporates into the TaN layers near the interfaces, appears particularly effective in enhancing energy dissipation during fracture, consistent with its pronounced increase in $J_n$. In contrast, silicon tends to remain confined to atomically thin interfacial enrichment layers (and diffusion into both adjacent layers, TiN and TaN), leading to substantial but slightly lower toughness gains, despite its superior hardness enhancement.

The present fracture toughness values place the investigated TiN/TaN superlattices in direct continuity with earlier benchmarking studies. Ref. [1] compiled hardness and fracture toughness data for undoped and B-doped TiN/TaN superlattices and demonstrated their favourable hardness–toughness balance compared to a wide range of ceramic coatings. However, the mechanistic origins of this behaviour remained unresolved. The present work establishes this missing link by directly correlating fracture resistance with interface structure, dopant distribution, elastic anisotropy, and strain accommodation at bilayer periods of ~6 nm.

These results also complement studies such as Ref. [22], where enhanced fracture toughness ($K_{IC}$ = 3.4 ± 0.5 MPa√m) was achieved at much larger bilayer periods (~83 nm) through stacking-fault-mediated deformation within thick TaN layers, driven by the intrinsic tendency of TaN toward hexagonal stacking [15]. In contrast, the present study demonstrates that comparable or higher fracture toughness can be achieved at nanoscale periodicities, where plastic deformation within individual layers is strongly suppressed. This identifies a fundamentally different toughening pathway governed by interface density, elastic contrast, and dopant-modified interfacial bonding rather than bulk-like deformation mechanisms.

Overall, these results demonstrate that TiN/TaN superlattices can achieve high fracture resistance without sacrificing nanoscale periodicity or hardness. Defect engineering through targeted interfacial doping emerges as a powerful and scalable strategy for simultaneously optimizing stiffness, hardness, and fracture toughness in ceramic superlattice coatings.

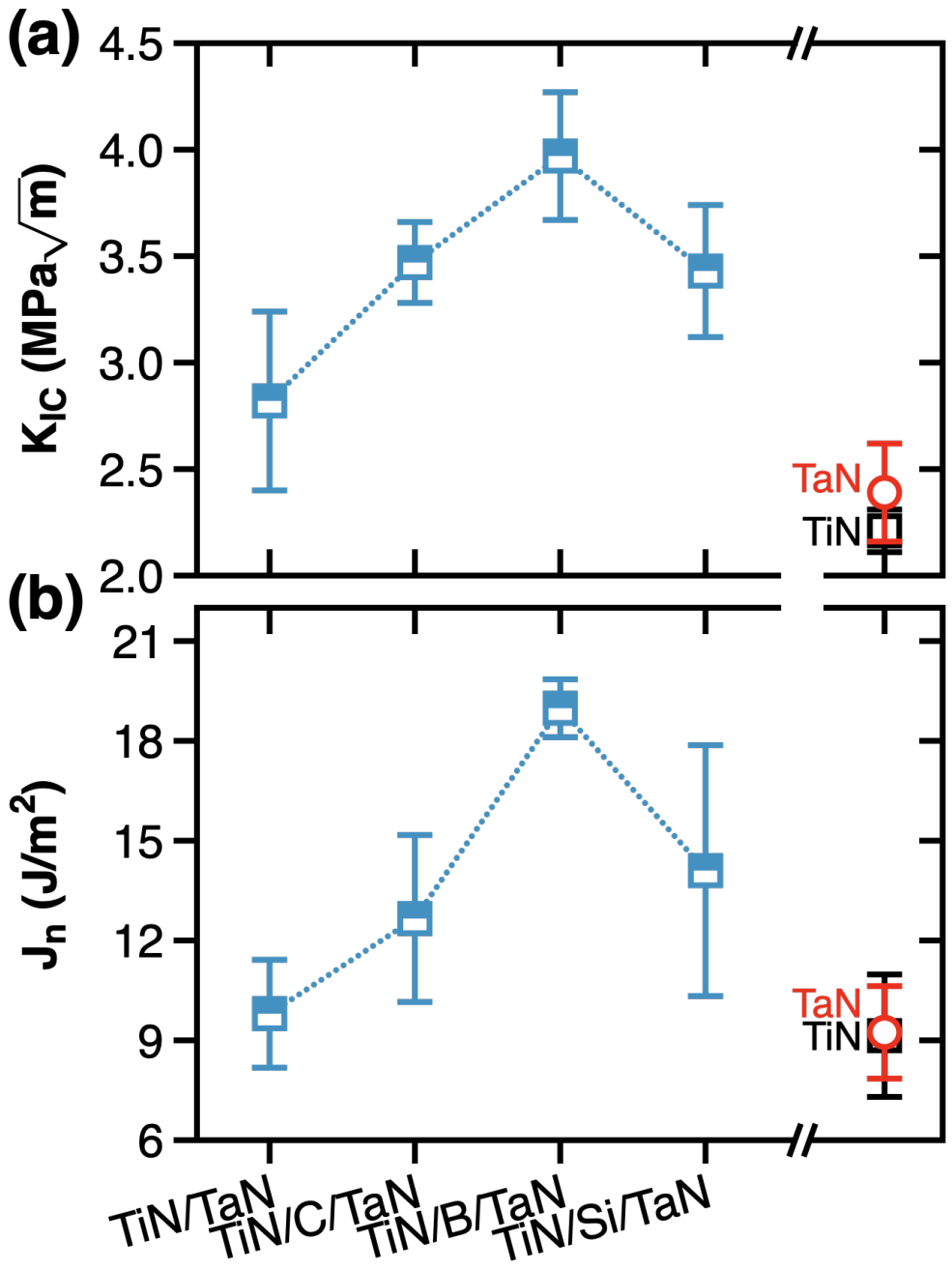

Figure 8: (a) Fracture toughness $K_{IC}$ and (b) integrated fracture energy $J_n$ of TiN/TaN, TiN/C/TaN, TiN/B/TaN, and TiN/Si/TaN superlattices. Values for monolithically grown TiN and TaN are included for comparison. Values are mean ± one standard deviation from n = 5–6 cantilevers per condition; error bars reflect this scatter. All data correspond to films grown on MgO(001) substrates.

## 4 Discussion

The present results place TiN/TaN superlattices within a mechanistically robust materials design framework for nitride superlattices with improved fracture toughness. Systematic DFT-based studies have shown that only a limited subset of nitride SLs can simultaneously achieve high hardness and fracture toughness through an optimized interplay of elastic anisotropy, coherency strain, and interface-controlled bonding [10]. Within this framework, TiN/TaN belongs to a top-performing group—including HfN/$Ta_{0.75}N$, TiN/$WN_{0.75}$, $MoN_{0.5}$/TaN, TiN/$Ta_{0.75}N$, and TiN/$MoN_{0.5}$—where the frequent occurrence of defect-stabilized sublayers, including $Ta_xN$, highlights their critical role in balancing stiffness and compliance at coherent interfaces.

ERDA analysis provides direct experimental support for this framework, revealing a pronounced nitrogen superstoichiometry in TaN layers (~55 at% N, with Ta/(Ta+Ti) of ~0.59 and the valid assumption of stoichiometric TiN), suggesting a high density of intrinsic Ta vacancies. Interfacial doping with B or Si shifts the composition closer to near-stoichiometric values (50–51 at% N, with Ta/(Ta+Ti) in the range of ~0.46 and 0.51), indicating a substantial modification of the defect landscape at both the layer and interface level. While Si is expected to increase interfacial cohesive strength and thereby promoting higher hardness [23], B may facilitate localized stacking-fault formation in the cubic nitrides. Within the TiN/TaN superlattice architecture, this combination of interface density and point-defect activity could activate defect-mediated, interface-triggered dislocation emission under load. Similar mechanisms have been observed in metastable TiN/AlN [59] and TiN/$MoN_{0.5}$ [54] SLs, where coherent interfaces and point-defect complexes act as efficient dislocation sources, enabling controlled microplasticity and enhanced fracture toughness.

### 4.1 Superlattice architecture and coherency-controlled strengthening

For the undoped TiN/TaN superlattice with a nominal bilayer period of ~6 nm, the measured hardness of ~34 GPa exceeds that of monolithic TiN (28 GPa) and TaN (31 GPa), demonstrating a clear superlattice strengthening effect beyond simple rule-of-mixtures behaviour. This response follows established superlattice hardening principles, in which coherency strain, elastic contrast, and interface-mediated dislocation mechanisms govern plastic deformation rather than volumetric averaging [1,14]. The close lattice match between TiN and defect-stabilized cubic $Ta_xN_y$ supports coherent or semi-coherent interfaces across several nanometres, consistent with the presence of well-defined XRD satellite reflections, and the cumulative 200 reflection observed in Figure 2.

At this bilayer period, the TiN/TaN system resides near the critical coherency thickness, where misfit dislocations may begin to form but do not fully relax the elastic strain. This regime maximizes resistance to dislocation glide while still permitting controlled, interface-mediated strain accommodation, a balance that is essential for achieving both high hardness and improved fracture resistance. In the undoped superlattice, the satellite reflections are relatively symmetric, indicating a nearly balanced strain distribution between the constituent layers.

Upon interfacial doping, the satellite reflections develop a pronounced intensity asymmetry, with negative-order satellites becoming markedly stronger. Given the larger lattice parameter and lower elastic stiffness of TaN compared to TiN, this evolution indicates that coherency strain and lattice distortions are increasingly accommodated within the TaN-based layers (as also shown through TEM and APT investigations, Figure 3 and Figure 4). The amplified asymmetry therefore reflects enhanced elastic and chemical contrast at otherwise coherent interfaces rather than a loss of epitaxy. This selective strain localization in TaN further reinforces the superlattice strengthening effect by maintaining high load-bearing stiffness in the TiN layers while enabling compliant, energy-dissipative deformation at the interfaces.

### 4.2 Dopant-dependent interface chemistry and controlled deviation from the baseline design space

Interfacial doping introduces an additional degree of freedom by modifying interface chemistry and local defect populations while preserving the coherency-controlled superlattice architecture established in the undoped system. ERDA and APT analyses reveal distinct dopant-dependent segregation behaviours: C and B diffuse predominantly into adjacent δ-TaN layers, whereas Si diffuses into both TiN and TaN and remains preferentially enriched at their interfaces. These segregation profiles directly govern the resulting structural coherence and mechanical response. The oxygen incorporation observed by APT is associated with the early stages of deposition and decreases substantially during subsequent growth, as confirmed by ERDA measurements of the steady-state coating composition. Therefore, although oxygen is present at the interfaces, its concentration does not correlate with the observed mechanical trends. The highest-hardness Si-doped architecture exhibits the lowest ERDA oxygen concentration, whereas the highest-toughness B-doped superlattice contains a different oxygen level. This indicates that the mechanical response is governed primarily by the intentionally introduced interface chemistry rather than uncontrolled oxygen incorporation.

Si uniquely enhances superlattice properties by remaining at the interfaces while simultaneously incorporating into both adjacent layers. This behaviour gives rise to the most pronounced superlattice characteristics, evidenced by sharper and more intense XRD satellite reflections and the highest hardness (~41 GPa, corresponding to ~22% improvement over the undoped superlattice). By strengthening interfacial chemical bonding, maintaining elastic contrast between the layers, and maximizing coherency stresses, Si drives the classical superlattice hardening mechanism to its full potential.

In contrast, C- and B-doped superlattices exhibit dopant distributions predominantly localized within the TaN layers. While this localization reduces the effectiveness of coherency-controlled strengthening compared to the Si-doped architecture, it introduces local lattice distortions and modifies vacancy populations within the TaN layers. These defect-mediated compliance variations promote energy dissipation during deformation and crack propagation, thereby enhancing toughness. Notably, the B-doped superlattice exhibits the highest measured fracture toughness (4.0 MPa√m) and $J_n$ (Fig. 8) despite a lower hardness than the Si-doped coating. Because the B-doped sample contains only ~0.5 at% C (ERDA) and the C-doped sample does not exhibit the highest $K_{IC}$, we attribute the toughness increase in the B-doped SL primarily to boron rather than to incidental carbon. This observation indicates that B doping establishes the most favourable balance between strengthening and toughness, suggesting that B-modified interfaces facilitate enhanced localized stress relaxation and crack-tip energy dissipation while retaining substantial strengthening. These findings demonstrate that interfacial dopants enable deliberate tuning of the hardness–toughness balance without disrupting the underlying superlattice framework.

### 4.3 Elastic anisotropy and interface-controlled fracture resistance

The experimentally observed increase in fracture toughness in the doped TiN/TaN superlattices correlates systematically with the measured nitrogen content, assuming that TiN layers remain stoichiometric and that variations in nitrogen concentration primarily reflect changes within the $Ta_xN_y$ sublayers. This interpretation is supported by DFT formation energy calculations (Figure S5 in the Supplementary Materials), which show that TiN/TaN interfaces destabilize upon introducing Ti or N vacancies into the TiN sublattice. In contrast, vacancies (both metallic and N) become energetically preferred when introduced into TaN layers, with up to 0.7 eV/at. stabilization relative to the stoichiometric TiN/TaN configuration. The progressive reduction in nitrogen content from 55 at% in the undoped TiN/TaN superlattice to 51 at% (Si), 51 at% (C), and 50 at% (B) therefore reflects an increasing concentration of vacancies within the TaN-based layers, likely corresponding to a gradual transition from $Ta_{0.75}N$ toward Schottky-defected $Ta_{0.875}N_{0.875}$.

First-principles calculations further reveal that vacancy incorporation in TaN causes pronounced elastic softening, reflected by reduced shear moduli and strongly positive Cauchy pressures, accompanied by lower $G/B$ ratios, particularly for Schottky-defected $Ta_{0.875}N_{0.875}$ (Table 2). While the reduction in $G/B$ qualitatively reflects the increased elastic compliance, it is not expected to quantitatively predict fracture toughness for these closely related nitride systems. Importantly, DFT predicts only modest variations in the intrinsic fracture toughness of TaN with increasing vacancy concentration, demonstrating that the experimentally observed $K_{IC}$ enhancement does not arise from a fundamentally tougher TaN constituent. Instead, the increased compliance of the TaN layers amplifies the elastic contrast within the superlattice, promoting more effective strain redistribution and crack-tip stress relaxation at coherent interfaces. This interpretation is corroborated by structural analysis of fully relaxed superlattice models (Figure S6 in the Supplementary Materials), which reveal that N vacancies—and even more prominently Ta vacancies—induce pronounced bond-length and bond-angle distortions within the TaN layers. These local deviations from ideal fcc symmetry generate a heterogeneous bonding network and localized strain fields that facilitate stress redistribution and increase the capacity for crack-energy dissipation.

However, the simultaneous increase in hardness and fracture toughness observed particularly for the B-doped superlattice indicates that elastic softening alone cannot account for the measured toughening response. Instead, the results suggest an optimized interplay between interface strengthening and localized defect-mediated energy dissipation. Additional support for dopant-induced local defect formation is provided by HRTEM and FFT analysis of the Si-doped superlattice, which reveals diffuse scattering and localized deviations from ideal lattice periodicity, indicative of increased lattice distortion associated with interfacial dopant incorporation (Fig. 3).

Similar behaviour has been reported for epitaxial TiN/(Cr,Al)N superlattices, where increased dislocation densities coincide with enhanced hardness and fracture resistance [37]. Estimating the associated defect strain energy using the classical dislocation line-energy approximation (~$0.5Gb^2$, with $G$ ≈100 GPa and $b$ ≈0.3 nm) suggests that the reported increase in dislocation density from $(2.4\pm0.6)\cdot10^{16}$ $1/m^2$ to $(4.7\pm0.9)\cdot10^{16}$ $1/m^2$ corresponds to an additional stored energy on the order of $10^8$ $J/m^3$. Although the present experiments do not directly quantify deformation-induced dislocation densities during crack propagation, the energetic estimate indicates that even moderate increases in defect density can provide a significant contribution to fracture-energy dissipation. When dissipated within nanometre-scale crack-tip process zones, where dislocation generation and interaction are expected to be most pronounced, this stored defect energy can contribute several $J/m^2$ to the fracture-energy balance. This estimate is consistent with the increased strain energy release rate reported for TiN/(Cr,Al)N superlattices [37] and with the enhanced energy dissipation observed here. Together with previous observations for TiN/MoN and TiN/AlN superlattices [54,59], these results support the concept that interface-mediated dislocation activity provides an additional pathway for energy dissipation during fracture. The B-doped superlattice, exhibiting the lowest nitrogen content and highest fracture toughness, therefore represents an optimized balance between elastic compliance, interface-mediated strengthening, and localized defect-assisted crack-tip energy dissipation, rather than a transition toward bulk plasticity or intrinsically weak bonding.

Among the five highest-performing nitride superlattices identified in previous DFT studies [10], all systems exhibit intermediate $G/B$ ratios together with non-zero Cauchy pressures, reflecting a common balance between stiffness and elastic compliance. Although neither parameter alone predicts fracture toughness, their consistent occurrence across these high-performing systems suggests that such an elastic response is favourable for achieving damage-tolerant superlattice architectures. $TiN/Ta_{0.75}N$ occupies a central position within this group, exhibiting elastic constants and fracture-related indicators comparable to $HfN/Ta_{0.75}N$ and $TiN/WN_{0.75}$. Although empirical fracture models underestimate the absolute $K_{IC}$ values, they correctly reproduce the relative ranking of TiN/TaN among the most fracture-resistant nitride superlattices, in excellent agreement with the present experiments.

Micromechanical bending experiments reveal linear-elastic behaviour up to catastrophic failure, with crack initiation occurring at FIB-defined pre-notches. The observed increases in fracture toughness ($K_{IC}$) and integrated fracture energy ($J_n$) therefore reflect intrinsic crack-growth resistance rather than macroscopic plasticity. Interface-controlled mechanisms dominate this behaviour: coherency strain fields and elastic mismatch promote crack deflection, misfit dislocations relax local stress concentrations, and chemically strengthened or compliance-tuned interfaces resist premature decohesion. Interfacial doping further refines these mechanisms by selectively tuning interface cohesion and local compliance without disrupting the superlattice framework. Consistent with DFT formation energy calculations (Figure S7 in the Supplementary Materials), smaller impurity atoms such as B, C, and O preferentially occupy N-vacancy sites, whereas larger atoms such as Si favour Ta or Ti vacancy positions, inducing additional local distortions beyond ideal fcc symmetry. Interestingly, C and O even reduce the defect formation energy at concentrations of ~1.7%, thereby stabilizing the interface. Through these combined effects, TiN/TaN superlattices achieve simultaneous enhancement of hardness and fracture resistance, consistent with design principles established for the most robust nitride superlattice systems.

At the present bilayer period (~6 nm), the superlattice remains too thin to sustain extended twinning or bulk-like stacking-fault-mediated deformation within TaN layers. Nevertheless, the intrinsic tendency of metastable fcc TaN (δ-TaN) to adopt hexagonal stacking (ε-TaN) remains mechanically relevant, lowering the barrier for partial dislocation emission at coherent or semi-coherent interfaces under high local stress [16–18]. Previous studies on TiN/AlN [59] and TiN/MoN [54] superlattices have demonstrated that interfaces in metastable NaCl-type architectures can act as effective dislocation sources, enabling localized plastic accommodation even when individual layers lack conventional plasticity. By analogy, the enhanced fracture toughness observed here likely originates from interface-mediated dislocation nucleation and localized defect accommodation, facilitated by coherency strain, elastic mismatch, and dopant-modified interfacial bonding. These mechanisms efficiently relax local stresses and dissipate energy during crack propagation while preserving high hardness.

Finally, the fracture toughness of the present TiN/TaN superlattices compares favourably with a broad range of advanced ceramic coatings, as benchmarked in Ref. [1]. Within this wider materials landscape, TiN/TaN-based architectures rank among the top-performing ceramic systems in terms of both absolute hardness and fracture toughness, as well as their combined hardness–toughness balance. Subsequent studies, e.g., Ref. [22], further identify TiN/TaN superlattices among the most fracture-resistant nitride coatings reported to date. The present results therefore establish TiN/TaN as a benchmark nitride superlattice system for achieving exceptional mechanical performance through interface-controlled design, demonstrating that tailored interfacial chemistry and coherency engineering can rival or surpass more compositionally complex coating concepts.

## 5 Summary and Conclusions

This work demonstrates how atomic-scale interface engineering controls the mechanical performance of TiN/TaN superlattice coatings and establishes mechanistic links between interface chemistry, defect structure, and fracture resistance. During growth, thin layers of Si, C, and B were intentionally introduced at the TiN/TaN interfaces to tailor their structural and mechanical properties, with ERDA confirming concentrations of 0.3, 2.1, and 0.1 at%, respectively. Comprehensive structural and chemical characterization confirms the formation of coherent cubic architectures with a bilayer period of ~6 nm and chemically well-defined interfaces. Distinct dopant-dependent segregation behaviours emerge: Si remains preferentially at the TiN/TaN interfaces while partially diffusing into both adjacent layers, whereas C and B diffuse predominantly into the TaN-based layers. These dopants modify local coherency strain, vacancy populations, and bonding configurations without disrupting the underlying superlattice architecture.

Mechanical testing reveals pronounced strengthening arising from the combined effects of superlattice structuring and targeted interfacial doping. Hardness increases from 28 GPa for TiN and 31 GPa for TaN to 34 GPa for undoped TiN/TaN superlattices and further to 41 GPa for Si-doped architectures, accompanied by an increase in indentation modulus from ~402 GPa to 425–473 GPa. These values substantially exceed rule-of-mixtures expectations and reflect strong interface strengthening, coherency strain hardening, and dopant-induced lattice distortions. Among all investigated coatings, Si-doped TiN/TaN exhibits the highest hardness, demonstrating that Si most effectively maximizes the classical superlattice hardening mechanism.

Microcantilever bending experiments reveal a concurrent and systematic improvement in fracture resistance. Fracture toughness increases from 2.2 MPa√m for TiN and 2.4 MPa√m for TaN to 2.8 MPa√m for the undoped TiN/TaN superlattice, in agreement with DFT predictions. Additional interfacial doping further elevates $K_{IC}$ to ~3.5 MPa√m (C), 4.0 MPa√m (B), and 3.4 MPa√m (Si). The B-doped superlattice exhibits the highest fracture toughness and the greatest energy dissipation during catastrophic fracture, demonstrating that B most effectively enhances crack-growth resistance while promoting additional fracture-energy dissipation.

First-principles calculations identify TiN as intrinsically more brittle than TaN, whereas vacancy-stabilized $Ta_xN_y$ exhibits greater elastic compliance, reflected by lower shear moduli, positive Cauchy pressures, and lower G/B ratios. While these elastic descriptors alone do not quantitatively predict fracture toughness, they consistently indicate an increased capacity for local shear accommodation. Vacancy-stabilized $Ta_xN_y$ does not possess substantially higher intrinsic fracture toughness; instead, its primary role is to increase elastic compliance and elastic contrast within the superlattice. However, the pronounced toughness enhancement observed in doped superlattices—particularly for B-doped architectures—cannot be explained by elastic mismatch alone. Rather, the combined experimental observations and first-principles calculations indicate that dopant-controlled interfacial effects, including modified coherency strain fields, altered local bonding environments, enhanced localized stress relaxation, and increased defect-mediated crack-tip energy dissipation, contribute significantly to fracture resistance.

Overall, the results reveal a clear dopant-dependent mechanical design principle: whereas Si doping maximizes hardness by promoting the strongest interfacial strengthening and coherency-controlled superlattice hardening, B doping provides the most favourable balance between strengthening and toughness by enabling enhanced crack-tip energy dissipation while preserving high hardness. Consequently, B-doped TiN/TaN achieves the most balanced overall mechanical performance, combining high hardness (~38 GPa), moderate elastic modulus (~425 GPa), and exceptional fracture toughness (~4.0 MPa√m). The contrasting behaviour of Si and B demonstrates that interface chemistry can be used to selectively optimize either maximum hardness or the balance between hardness and fracture resistance.

This study establishes a mechanistic framework in which interface chemistry, elastic contrast, and defect-assisted energy dissipation act synergistically to overcome the classical hardness–toughness trade-off in ceramic nitride coatings. By combining coherency strain, elastic contrast, and dopant-engineered interfacial chemistry,

TiN/TaN superlattices achieve simultaneous enhancement of strength and damage tolerance through interface-dominated energy dissipation mechanisms while preserving high load-bearing capacity. These findings identify chemically engineered coherent superlattices as a robust and generalizable materials-design strategy for advanced ceramic coatings operating under extreme mechanical and thermal conditions.

## Declaration of Generative AI and AI-assisted technologies in the writing process

During the preparation of this work the authors used DeepL (www. deepl.com) in order to improve the readability of certain sentences. After using this tool, the authors reviewed and edited the content as needed and take full responsibility for the content of the published article.

## Declaration of interests

The authors declare that they have no known competing or financial interests, or personal relationships, that could have influenced the work reported in this paper.

## Data Availability

The data that supports the findings of this study are available from the authors on reasonable request.

## Acknowledgements

The authors gratefully acknowledge support from the Austrian Science Fund (FWF) under project 10.55776/PAT4425523 (Interfaces). We acknowledge the use of the USTEM electron microscopy facility and the X-ray Center (XRC) at TU Wien. Parts of the computational work were performed using the Vienna Scientific Cluster (VSC). ZCG gratefully acknowledges support from the National Natural Science Foundation of China (Grant No. 52401083). The films were grown by Helena Luise Hazeu and Marian Koller during their Bachelor theses at TU Wien in 2020. MH and JMS are grateful for financial support from Deutsche Forschungsgemeinschaft (DFG) within the project 515702322 (HA 9139/1-1, SCHN 735/50-1). Accelerator operation at Uppsala University has been supported by the Swedish Research Council (VR- RFI) within grant agreement #2019-00191. All sputtering targets were provided by Plansee Composite Materials GmbH. Open Access funding was provided by the TU Wien Bibliothek.

## References

1. P.H. Mayrhofer, H. Clemens, F.D. Fischer Materials-science based guidelines to develop robust hard thin film materials, Prog. Mater. Sci. 146 (2024) 101323 1–57. https://doi.org/10.1016/j.pmatsci.2024.02.101323
2. C. Mitterer, P. Mayrhofer, Some Materials Science Aspects of PVD Hard Coatings, Advances in Solid State Physics, in: B. Kramer (Eds.), Springer Berlin / Heidelberg, 2001, pp. 263-274.
3. A. Baptista, F. Silva, J. Porteiro, J. Míguez, G. Pinto, Sputtering Physical Vapour Deposition (PVD) Coatings: A Critical Review on Process Improvement and Market Trend Demands, Coatings 8 (2018). 10.3390/coatings8110402
4. K. Bobzin, High-performance coatings for cutting tools, CIRP J. Manuf. Sci. Technol. 18 (2017) 1-9. http://dx.doi.org/10.1016/j.cirpj.2016.11.004
5. R. Hahn, M. Bartosik, R. Soler, C. Kirchlechner, G. Dehm, P.H. Mayrhofer, Superlattice effect for enhanced fracture toughness of hard coatings, Scr. Mater. 124 (2016) 67-70. https://doi.org/10.1016/j.scriptamat.2016.06.030
6. A. Wagner, D. Holec, P.H. Mayrhofer, M. Bartosik, Enhanced fracture toughness in ceramic superlattice thin films: On the role of coherency stresses and misfit dislocations, Mater. Des. 202 (2021) 109517. https://doi.org/10.1016/j.matdes.2021.109517
7. Chu X, Barnett SA. Model of superlattice yield stress and hardness enhancements. J Appl Phys 1995;77:4403–11.
8. Helmersson U, Todorova S, Barnett SA, Sundgren J-E, Markert LC, Greene JE. Growth of single-crystal TiN/VN strained-layer superlattices with extremely high mechanical hardness. J Appl Phys 1987;62:481–484.
9. Koehler JS. Attempt to design a strong solid. Phys Rev B 1970;2:547–51
10. N. Koutná, A. Brenner, D. Holec, P.H. Mayrhofer, High-throughput first-principles search for ceramic superlattices with improved ductility and fracture resistance, Acta Mater. 206 (2021) 116615.

https://doi.org/10.1016/j.actamat.2020.116615

11. J. Buchinger, N. Koutná, Z. Chen, Z. Zhang, P.H. Mayrhofer, D. Holec, M. Bartosik, Toughness enhancement in TiN/WN superlattice thin films, Acta Mater. 172 (2019) 18-29. https://doi.org/10.1016/j.actamat.2019.04.028
12. Rainer Hahn, Nikola Koutná, Tomasz Wójcik, Anton Davydok, Szilárd Kolozsvári, Christina Krywka, David Holec, Matthias Bartosik, Paul H. Mayrhofer, Mechanistic study of superlattice-enabled high toughness and hardness in MoN/TaN coatings. Commun Mater **1**, 62 (2020). https://doi.org/10.1038/s43246-020-00064-4
13. Z. Gao, J. Buchinger, N. Koutná, T. Wojcik, R. Hahn, P.H. Mayrhofer, Ab initio supported development of TiN/MoN superlattice thin films with improved hardness and toughness, Acta Mater. 231 (2022) 117871. https://doi.org/10.1016/j.actamat.2022.117871
14. Sen Yang, Tao Guo, Xueyan Yan, Kewei Gao, Jingwen Qiu, Xiaolu Pang, Nanotwinned transition metal nitride coating with excellent thermal stability, Acta Materialia, Volume 267, 2024, 119743, ISSN 1359-6454, https://doi.org/10.1016/j.actamat.2024.119743.
15. Nitin Patel, Shanling Wang, Aharon Inspektor, Paul A. Salvador, Secondary hardness enhancement in large period TiN/TaN superlattices, Surface and Coatings Technology, Volume 254, 2014, Pages 21-27, ISSN 0257-8972, https://doi.org/10.1016/j.surfcoat.2014.05.030
16. C. Stampfl and A. J. Freeman, Stable and metastable structures of the multiphase tantalum nitride system, Phys. Rev. B 71 (2005) 024111 10.1103/PhysRevB.71.024111
17. Michael Grumski, Pratik P. Dholabhai, James B. Adams, Ab initio study of the stable phases of 1:1 tantalum nitride, Acta Materialia 61 (2013) 3799-3807 10.1016/j.actamat.2013.03.018
18. N. Koutná, D. Holec, O. Svoboda, F.F. Klimashin, P.H. Mayrhofer, Point defects

stabilise cubic Mo-N and Ta-N, J. Phys. D: Appl. Phys. 49 (2016) 375303 doi.org/10.1088/0022-3727/49/37/375303

19. W. Ensinger, M. Kiuchi, M. Satou, Low-temperature formation of metastable cubic tantalum nitride by metal condensation under ion irradiation, J. Appl. Phys. 77 (1995) 6630–6635 https://doi.org/10.1063/1.359073
20. C.-S. Shin, Y.-W. Kim, N. Hellgren, D. Gall, I. Petrov, J. E. Greene, Epitaxial growth of metastable δ-TaN layers on MgO(001) using low-energy, high-flux ion irradiation during ultrahigh vacuum reactive magnetron sputtering, J. Vac. Sci. Technol. A 20 (2002) 2007–2017 10.1116/1.1513639
21. Yong Huang, Zhuo Chen, Antonia Wagner, Christian Mitterer, Kexing Song, Zaoli Zhang, High density of stacking faults strengthened TaN/TiN multilayer, Acta Materialia, Volume 255, 2023, 119027, ISSN 1359-6454, https://doi.org/10.1016/j.actamat.2023.119027.
22. Yong Huang, Zhuo Chen, Michael Meindlhumer, Rainer Hahn, David Holec, Thomas Leiner, Verena Maier-Kiener, Yonghui Zheng, Zequn Zhang, Lukas Hatzenbichler, Helmut Riedl, Christian Mitterer, Zaoli Zhang, Harvesting superior intrinsic plasticity in nitride ceramics with negative stacking fault energy, Acta Materialia, Volume 286, 2025, 120774, ISSN 1359-6454, https://doi.org/10.1016/j.actamat.2025.120774.
23. Söderberg H, Odén M, Flink A, et al. Growth and characterization of TiN/SiN(001) superlattice films. Journal of Materials Research. 2007;22(11):3255-3264. doi:10.1557/JMR.2007.0412
24. P. Ström and D. Primetzhofer, Ion beam tools for nondestructive in-situ and in-operando composition analysis and modification of materials at the Tandem Laboratory in Uppsala, 2022 JINST 17 P04011 https://doi.org/10.1088/1748-0221/17/04/P04011
25. Yanwen Zhang, Harry J. Whitlow, Thomas Winzell, Ian F. Bubb, Timo Sajavaara, Kai Arstila, Juhani Keinonen, Detection efficiency of time-of-flight energy elastic recoil detection analysis systems, Nuclear Instruments and Methods in Physics Research Section B: Beam Interactions with Materials and Atoms 149 (1999) 477–489, https://doi.org/10.1016/S0168-583X(98)00963-X

26. Petter Ström, Per Petersson, Marek Rubel, Göran Possnert; A combined segmented anode gas ionization chamber and time-of-flight detector for heavy ion elastic recoil detection analysis. Rev. Sci. Instrum. 1 October 2016; 87 (10): 103303. https://doi.org/10.1063/1.4963709
27. to Baben, M., Hans, M., Primetzhofer, D., Evertz, S., Ruess, H., & Schneider, J. M. (2017). Unprecedented thermal stability of inherently metastable titanium aluminum nitride by point defect engineering. Materials Research Letters, 5(3), 158–169. https://doi.org/10.1080/21663831.2016.1233914
28. M. A. Sortica1, V. Paneta, B. Bruckner, S. Lohmann, M. Hans, T. Nyberg, P. Bauer, and D. Primetzhofer, Electronic energy-loss mechanisms for H, He, and Ne in TiN, Phys. Rev. A 96 (2017) 032703 https://doi.org/10.1103/PhysRevA.96.032703
29. Horton, M.K., Huck, P., Yang, R.X. et al. Accelerated data-driven materials science with the Materials Project. Nat. Mater. 24, 1522–1532 (2025). https://doi.org/10.1038/s41563-025-02272-0 (TiN: https://doi.org/10.17188/1208488; TaN: https://doi.org/10.17188/1200491)
30. Wang, Y. & Nastasi, M. Handbook of Modern Ion Beam Materials Analysis (MRS, Materials Research Soc, 2010)
31. R. Saha, W.D. Nix, Effects of the substrate on the determination of thin film mechanical properties by nanoindentation, Acta Mater. 50 (2002) 23-38. https://doi.org/10.1016/S1359-6454(01)00328-7
32. A.C. Fischer-Cripps, Nanoindentation, Springer, 2011, https://doi.org/10.1007/978-1-4419-9872-9.
33. F.F. Klimashin, P.H. Mayrhofer, Ab initio-guided development of super-hard Mo–Al–Cr–N coatings, Scripta Materialia 140 (2017) 27–30,

https://doi.org/10.1016/j.scriptamat.2017.06.052.

34. W.C. Oliver, G.M. Pharr, An Improved Technique for Determining Hardness and Elastic-Modulus Using Load and Displacement Sensing Indentation Experiments, Journal of Materials Research 7(6) (1992) 1564–1583, https://doi.org/10.1557/Jmr.1992.1564.
35. A.C. Fischer-Cripps, Critical review of analysis and interpretation of nanoindentation test data, Surface and Coatings Technology 200(14–15) (2006) 4153–4165, https://doi.org/10.1016/j.surfcoat.2005.03.018.
36. S. Brinckmann, K. Matoy, C. Kirchlechner, G. Dehm, On the influence of microcantilever pre-crack geometries on the apparent fracture toughness of brittle materials, Acta Mater. 136 (2017) 281-287. https://doi.org/10.1016/j.actamat.2017.07.014
37. J. Buchinger, A. Wagner, Z. Chen, Z.L. Zhang, D. Holec, P.H. Mayrhofer, M. Bartosik, Fracture toughness trends of modulus-matched TiN/(Cr,Al)N thin film superlattices, Acta Mater. 202 (2021) 376-386. https://doi.org/10.1016/j.actamat.2020.10.068
38. G. Kresse and J. Furthmüller, Physical Review B 54, 11169 (1996).
39. G. Kresse and D. Joubert, Physical Review B 59, 1758 (1999).
40. W. Kohn and L. J. Sham, Physical Review 140, A1133 (1965).
41. J. P. Perdew, K. Burke, and M. Ernzerhof, Physical Review Letters 77, 3865 (1996).
42. H. J. Monkhorst and J. D. Pack, Physical review B 13, 5188 (1976).
43. D. Gehringer, M. Friák, and D. Holec, Computer physics communications 286, 108664 (2023).
44. Y. Le Page and P. Saxe, Phys. Rev. B: Condens. Matter 65, 104104 (2002).
45. R. Yu, J. Zhu, and H. Ye, Comput. Phys. Commun. 181, 671 (2010).
46. D. G. Sangiovanni, V. Chirita, and L. Hultman, Physical Review B 81, 104107 (2010).

47. H. Niu, X.-Q. Chen, P. Liu, W. Xing, X. Cheng, D. Li, and Y. Li, Scientific reports 2, 1 (2012).
48. P. Lazar, J. Redinger, and R. Podloucky, Phys. Rev. B: Condens. Matter 76, 174112 (2007).
49. P. Řehák, M. Černý, and D. Holec, Surf. Coat. Technol. 325, 410 (2017).
50. M. Bielawski and K. Chen, Journal of Engineering for Gas Turbines and Power 133 (2010).
51. C.M. Koller, H. Marihart, H. Bolvardi, S. Kolozsvári, P.H. Mayrhofer, Structure, phase evolution, and mechanical properties of DC, pulsed DC, and high power impulse magnetron sputtered Ta–N films, Surf. Coat. Technol. 347 (2018) 304-312. https://doi.org/10.1016/j.surfcoat.2018.05.003
52. A. Zaman, E. I. Meletis, Microstructure and Mechanical Properties of TaN Thin Films Prepared by Reactive Magnetron Sputtering, Coatings, 7 (2017), 209, https://doi.org/10.3390/coatings7120209
53. T Riekkinen, J. Molarius; T. Laurila; A. Nurmela; I. Suni, J.K. Kivilahti, Reactive sputter deposition and properties of TaxN thin films. Microelectron. Eng. 2002, 64, 289–297.
54. Z. Chen, Y. Huang, Z. Gao, Y. Zheng, P.H. Mayrhofer, Z. Zhang, Direct observation of Schottky-vacancy clusters and their mechanical response in MoN/TiN superlattice, Acta Mat. 283 (2025) 120551 1–15 doi.org/10.1016/j.actamat.2024.120551
55. Yu-Ping Chien, Stanislav Mráz, Matej Fekete, Marcus Hans, Daniel Primetzhofer, Szilárd Kolozsvári, Peter Polcik, Jochen M. Schneider, Deviations between film and target compositions induced by backscattered Ar during sputtering from M2-Al-C (M = Cr, Zr, and Hf) composite targets, Surface and Coatings Technology 446 (2022) 128764 https://doi.org/10.1016/j.surfcoat.2022.128764
56. K. Matoy, H. Schönherr, T. Detzel, T. Schöberl, R. Pippan, C. Motz, G. Dehm,

A comparative micro-cantilever study of the mechanical behavior of silicon based passivation films, Thin Solid Films 518 (2009) 247-256. https://doi.org/10.1016/j.tsf.2009.07.143

57. D. Di Maio, S.G. Roberts, Measuring fracture toughness of coatings using focused-ion-beam-machined microbeams, Journal of Materials Research 20 (2005) 299-302. https://doi.org/10.1557/JMR.2005.0048
58. A.T. Zehnder, Fracture Mechanics, Springer Science+Business Media (2012).
59. Z. Chen, Y. Zheng, L. Löfler, M. Bartosik, G.K. Nayak, O. Renk, D. Holec, P.H. Mayrhofer, Z. Zhang, Atomic insights on intermixing of nanoscale nitride multilayer triggered by nanoindentation, Acta Mat. 214 (2021) 117004 1–11 doi.org/10.1016/j.actamat.2021.117004